%% file: paper.tex
\documentclass[letterpaper,twocolumn,10pt]{article}
\usepackage{usenix2019_v3}
\pdfoutput=1

\usepackage{latexsym}      
\usepackage{mathtools}	   
\usepackage{xspace}   
\usepackage{booktabs}
\usepackage{epsfig}
\usepackage{xcolor}   
\usepackage{cite}     
\usepackage{tabularx} 
\usepackage{pifont}
\usepackage[font=small,labelfont=bf]{caption}
\usepackage{balance}
\usepackage{times}
\usepackage{multirow}
\usepackage{graphicx}
\usepackage{subcaption}
\usepackage{enumitem}
\usepackage{comment}
\usepackage{algorithm}
\usepackage{algpseudocode}
\usepackage{listings}

\usepackage{csvsimple}
\usepackage{siunitx}
\usepackage[utf8]{inputenc}
\usepackage[title]{appendix}


\renewcommand{\paragraph}[1]{\vspace{5pt}\noindent\textbf{#1:}}

\begin{document}
\pagestyle{empty}

\date{}
\title{Detecting and Characterizing Lateral Phishing at Scale}

\def\icsi{\raisebox{6pt}{\small $\oint$}}
\def\ucb{\raisebox{6pt}{\small $\dagger$}}
\def\ucsd{\raisebox{6pt}{\small $\star$}} 
\def\barracuda{\raisebox{6pt}{\small $\circ$}}
\def\columbia{\raisebox{6pt}{\small $\psi$}}

\newcommand{\authspace}{\hspace{3mm}}
\newcommand{\titlespace}{\hspace{3mm}}

\author{
  Grant Ho\ucb\barracuda\authspace
  Asaf Cidon\barracuda\columbia\authspace
  Lior Gavish\barracuda\authspace
  Marco Schweighauser\barracuda
  \\
  Vern Paxson\ucb\icsi\authspace
  Stefan Savage\ucsd\authspace
  Geoffrey M. Voelker\ucsd\authspace
  David Wagner\ucb
\vspace{.2cm}\\
  {\small\barracuda{Barracuda Networks}}\titlespace
  {\small\ucb{UC Berkeley}}\titlespace
  {\small\ucsd{UC San Diego}}\titlespace
  {\small\columbia{Columbia University}}\titlespace
  {\small\icsi{International Computer Science Institute}}
}

\input{macros}
\maketitle

\input{abstract}
\input{intro}
\input{background}
\input{dataset}
\input{design}
\input{evaluation}
\input{characterization}
\input{conclusion}
\input{acknowledgements}

\bibliographystyle{plain}
\bibliography{references}
\input{appendix}

\end{document}

%% file: macros.tex
\newcommand{\cudaFullName}{{Barracuda Networks}\xspace}
\newcommand{\cuda}{{Barracuda}\xspace}

\newcommand{\eg}{e.g.,\xspace}
\newcommand{\ie}{i.e.,\xspace}
\newcommand{\trainingOrgs}{exploratory organizations\xspace}
\newcommand{\testOrgs}{test organizations\xspace}

\newcommand{\featurefuzzyphish}{\emph{fuzzy phish similarity score}\xspace}
\newcommand{\featurelocalrep}{\emph{local URL reputation}\xspace}
\newcommand{\featureglobalrep}{\emph{global URL reputation}\xspace}
\newcommand{\featuretemplatematch}{\emph{template similarity score}\xspace}
\newcommand{\featureRecipSimilarity}{recipient likelihood\xspace}
\newcommand{\detectorFuzzyPhish}{Fuzzy Phish Detector\xspace}
\newcommand{\detectorTemplate}{Template Detector\xspace}
\newcommand{\detectorViral}{Viral Detector\xspace}

\newcommand{\trainingMonths}{April--June 2018\xspace}
\newcommand{\evalMonths}{July--October 2018\xspace}
\newcommand{\numMonths}{7\xspace}

\newcommand{\totalOrgs}{92\xspace}
\newcommand{\numRandomOrgs}{69\xspace}
\newcommand{\numTrainingOrgs}{52\xspace}
\newcommand{\numTestOrgs}{40\xspace}

\newcommand{\numEmails}{113,083,695\xspace}
\newcommand{\numMailboxesTrainingOrgs}{89,267\xspace}  
\newcommand{\numMailboxesTestOrgs}{138,752\xspace}
\newcommand{\numEmailsTrainingOrgs}{60,143,337\xspace}  
\newcommand{\numEmailsTestOrgs}{52,940,358\xspace}  
\newcommand{\numEmailsTrainDataset}{25,670,264\xspace}  
\newcommand{\numEmailsTestDataset}{87,413,431\xspace}  

\newcommand{\numPhishEmailsFoundByDetectors}{1,377\xspace}
\newcommand{\numPhishIncidentsFoundByDetectors}{161\xspace}
\newcommand{\numPhishIncidentsFoundOnlyByTextDetectors}{two\xspace}   
\newcommand{\numAllPhishIncidentsFoundOnlyByTextDetectors}{2\xspace}

\newcommand{\numReportedPhishEmails}{1,694\xspace}  
\newcommand{\numIncidentsReported}{101\xspace}
\newcommand{\numPhishOrgsWithReported}{27\xspace}

\newcommand{\recallAggregateTest}{87.3\%\xspace} 
\newcommand{\fpPercentAggregateTest}{0.00036\%\xspace} 
\newcommand{\aggregateDetectorNumBenignPerFP}{277,000\xspace} 

\newcommand{\numPhishEmails}{1,902\xspace}
\newcommand{\numIncidents}{180\xspace}
\newcommand{\numAttachPhishEmails}{40\xspace}
\newcommand{\numAttachIncidents}{12\xspace}
\newcommand{\numAttachATO}{12\xspace}
\newcommand{\numURLPhishEmails}{1,862\xspace}
\newcommand{\numURLIncidents}{168\xspace}  

\newcommand{\numTrainIncidents}{70\xspace}
\newcommand{\numTestIncidents}{110\xspace}
\newcommand{\numReportedTrainIncidents}{40\xspace}
\newcommand{\numTrainIncidentsTextFoundFN}{2\xspace}
\newcommand{\numReportedTestIncidents}{61\xspace}
\newcommand{\numNewlyDiscoveredIncidents}{77\xspace}
\newcommand{\numNewlyDiscoveredTestIncidents}{49\xspace}

\newcommand{\numHijackedAccounts}{154\xspace}
\newcommand{\numPhishOrgs}{33\xspace}
\newcommand{\numOrgsWithSingleATO}{13\xspace}
\newcommand{\numOrgsWithMultiATO}{20\xspace}
\newcommand{\numPhishTrainingOrgs}{14\xspace}
\newcommand{\numPhishTestOrgs}{19\xspace}

\newcommand{\numPhishOrgsAPrioriWeKnew}{23\xspace}
\newcommand{\numPhishOrgsFromSampling}{10\xspace}
\newcommand{\percentRandomOrgsWithPhish}{14\%\xspace}  

\newcommand{\numSuccessfulATOs}{17\xspace}
\newcommand{\numSuccessfulIncidents}{17\xspace}
\newcommand{\numOrgsWithSuccessfulATOs}{10\xspace}
\newcommand{\numInfecteeATOs}{23\xspace}
\newcommand{\percentSuccessfulATO}{11\%\xspace}  
\newcommand{\numEmployeeRecipForSuccessfulInfectee}{542\xspace}
\newcommand{\minNumEmployeeRecipForSuccessfulInfectee}{26\xspace}

\newcommand{\numPhishRecipAll}{101,276\xspace}
\newcommand{\numPhishRecipExternal}{59,536\xspace}
\newcommand{\numPhishRecipEmployee}{41,740\xspace}
\newcommand{\numPhishRecipOrgsAll}{16,293\xspace}
\newcommand{\numPhishRecipOrgsExternal}{16,260\xspace}

\newcommand{\numATOsManyEmails}{10\xspace}
\newcommand{\numATOsLargeRecipEmails}{131\xspace}
\newcommand{\percentATOsTotalRecipLarge}{94\%\xspace}  

\newcommand{\atoRecipAgnostic}{Account-agnostic\xspace}
\newcommand{\atoOrgWide}{Organization-wide\xspace}
\newcommand{\atoContactBook}{Contact-book\xspace}
\newcommand{\atoTargetedRecip}{Targeted-recipient\xspace}
\newcommand{\atoLateralOrg}{Lateral-organization\xspace}

\newcommand{\numATOsRecipAgnostic}{63\xspace}
\newcommand{\percentATOsRecipAgnostic}{40.9\%\xspace}
\newcommand{\numATOsNoEmployeeRecip}{37\xspace}
\newcommand{\percentATOsNoEmployeeRecip}{24\%\xspace}
\newcommand{\numATOsNoEmployeeRecipAndAllPersonalDomains}{2\xspace}
\newcommand{\numATOsNoEmployeeRecipAndFewTotalDomains}{4\xspace}
\newcommand{\numATOsNoEmployeeRecipAndManyTotalDomains}{33\xspace}
\newcommand{\percentMaxContactBookOverlapRecipAgnostic}{17\%\xspace}
\newcommand{\numATOsRecipAgnosticNoEmployeeCriteria}{35\xspace}
\newcommand{\numATOsMajorityExternalRecipAndManyTotalDomains}{63\xspace}
\newcommand{\numATOsRecipAgonisticSecondVersion}{54\xspace}

\newcommand{\percentMaxContactBookOverlapOrgWideAgnostic}{11\%\xspace}
\newcommand{\minNumRecipForOrgWideAttackersGeqHalfRecip}{1,300\xspace}
\newcommand{\numATOsOrgWideGeqHalfOrgRecip}{16\xspace}
\newcommand{\numATOsNearlyAllEmployeeRecip}{36\xspace}
\newcommand{\percentATOsNearlyAllEmployeeRecip}{23.3\%\xspace}  
\newcommand{\numATOsOrgWideNearlyAllEmployeeRecip}{29\xspace}
\newcommand{\numATOsOrgWide}{39\xspace}  
\newcommand{\percentATOsOrgWide}{25.3\%\xspace}  

\newcommand{\numATOsLateralOrg}{2\xspace}

\newcommand{\numATOPotentialTargetedRecip}{50\xspace}
\newcommand{\numATOTargetedRecip}{44\xspace}
\newcommand{\percentATOTargetedRecipOfPotential}{88\xspace}
\newcommand{\percentATOTargetedRecipOverall}{28.6\%\xspace}
\newcommand{\numATOUnknownRecipTargeting}{6\xspace}

\newcommand{\numIncidentsContentRecipNamed}{3\xspace}
\newcommand{\percentIncidentsContentRecipNamed}{1.7\%\xspace} 
\newcommand{\numIncidentsContentNonTargeted}{167\xspace}
\newcommand{\percentIncidentsContentNonTargeted}{92.7\%\xspace} 
\newcommand{\numIncidentsContentTargeted}{13\xspace}
\newcommand{\percentIncidentsTargetedMsg}{7\%\xspace}
\newcommand{\numATOsContentTargeted}{13\xspace}

\newcommand{\numOrgsWithInteractiveATO}{15\xspace}
\newcommand{\numInteractiveATO}{27\xspace}
\newcommand{\numATOsWithInquiries}{107\xspace}
\newcommand{\numATOsWithNoInquiries}{47\xspace}
\newcommand{\percentInteractiveATO}{25\%\xspace}  
\newcommand{\numInteractiveAndTargetedContentATO}{1\xspace}
\newcommand{\numInteractiveAndEnterpriseContentATO}{7\xspace}

\newcommand{\percentIncidentsWorkDay}{98\%\xspace} 
\newcommand{\percentIncidentsFirstHalfOfWeek}{67\%\xspace} 
\newcommand{\numIncidentsActiveATO}{165\xspace} 
\newcommand{\numIncidentsQuiescentATO}{15\xspace} 
\newcommand{\numIncidentsOutsideNormalTime}{18\xspace} 
\newcommand{\numIncidentsNormalTime}{147\xspace} 
\newcommand{\percentIncidentsNormalTime}{81.7\%\xspace} 


\newcommand{\numStealthyATO}{30\xspace}
\newcommand{\numOrgsWithStealthyATO}{16\xspace}
\newcommand{\numStealthyAndInteractiveATO}{9\xspace}
\newcommand{\numStealthyNonInteractiveATO}{21\xspace}

\newcommand{\numAnySophisticationATO}{48\xspace}   
\newcommand{\percentSophisticatedATO}{31\%\xspace}   

%% file: abstract.tex
\begin{abstract}

We present the first large-scale characterization of lateral phishing attacks,
based on a dataset of 113 million employee-sent emails from \totalOrgs enterprise organizations.
In a lateral phishing attack,
adversaries leverage a compromised enterprise account to send phishing emails to other users,
benefitting from both the implicit trust and the information in the hijacked user's account.
We develop a classifier that
finds hundreds of real-world lateral phishing emails,
while generating
under four false positives per every one-million employee-sent emails.
Drawing on the attacks we detect,
as well as a corpus of user-reported incidents,
we quantify the scale of lateral phishing,
identify several thematic content and recipient targeting strategies that attackers follow,
illuminate two types of sophisticated behaviors that attackers exhibit,
and estimate the success rate of these attacks.
Collectively, these results expand our mental models of the `enterprise attacker'
and shed light on the current state of enterprise phishing attacks.
\end{abstract}

%% file: intro.tex
\section{Introduction}

For over a decade, the security community has explored a myriad of defenses against phishing attacks.
Yet despite this long line of work, modern-day attackers routinely and successfully use phishing attacks to compromise government systems, political figures, and companies spanning every economic sector.
Growing in prominence each year,
this genre of attacks has risen to the level of government attention,
with the FBI estimating \$12.5 billion in financial losses worldwide from 78,617 reported incidents between October 2013 to May 2018~\cite{eac-fbi-report},
and the US Secretary of Homeland Security declaring that phishing represents ``the most devastating attacks by the most sophisticated attackers''~\cite{homelandsecurity}.

By and large, the high-profile coverage around targeted spearphishing attacks against
major entities, such as Google, RSA, and the Democratic National Committee,
has captured and shaped our mental models of enterprise phishing attacks~\cite{presidentialEmailLeaks, rsabreach2, highprofileSpearphish}.
In these newsworthy instances, as well as many of the targeted spearphishing incidents discussed in the academic literature~\cite{le2014look, le2017broad, marczak2014governments},
the attacks come from external accounts, created by nation-state adversaries who cleverly craft or spoof the phishing account’s name and email address to resemble a known and legitimate user.
However, in recent years, work from both industry~\cite{barracudaATO, ftOnlineLateralPhish, chainPhishing}
and academia~\cite{ho2017detecting, bursztein2014handcrafted, onaolapo2016happens, stringhini2015ain} has pointed
to the emergence and growth of \emph{lateral phishing} attacks:
a new form of phishing that targets a diverse range of organizations and has already incurred billions of dollars in financial harm~\cite{eac-fbi-report}.
In a lateral phishing attack, an adversary uses a compromised enterprise account to send phishing emails to a new set of recipients.
This attack proves particularly insidious because the attacker automatically benefits from the implicit trust in the hijacked account:
trust from both human recipients and conventional email protection systems.

Although recent work~\cite{ho2017detecting, hu2016detecting, gascon2018reading, stringhini2015ain, duman2016emailprofiler}
presents several ideas for detecting lateral phishing,
these prior methods either require that organizations possess sophisticated network monitoring infrastructure, or they
produce too many false positives for practical usage.
Moreover, no prior work has characterized this attack at a large, generalizable scale.
For example, one of the most comprehensive related work uses a multi-year dataset from one organization, which only contains two lateral phishing attacks~\cite{ho2017detecting}.
This state of affairs leaves many important questions unanswered:
How should we think about this class of phishing with respect to its scale, sophistication, and success?
Do attackers follow thematic strategies, and can these common behaviors fuel new or improved defenses?
How are attackers capitalizing on the information within the hijacked accounts,
and what does their behavior say about the state and trajectory of enterprise phishing attacks?

In this joint work between academia and \cudaFullName
we take a first step towards answering these open questions and understanding \emph{lateral phishing} at scale.
This paper seeks to both explore avenues for practical defenses against this burgeoning threat
and develop accurate mental models for the state of these phishing attacks in the wild.

First, we present a new classifier for detecting URL-based lateral phishing emails and evaluate our approach on a dataset of 113~million emails, spanning \totalOrgs enterprise organizations.
While the dynamic churn and dissimilarity in content across phishing emails proves challenging,
our approach can detect \recallAggregateTest of attacks in our dataset,
while generating less than 4 false positives per every 1,000,000 employee-sent emails.

Second, combining the attacks we detect with a corpus of user-reported lateral phishing attacks,
we conduct the first large-scale characterization of lateral phishing in real-world organizations.
Our analysis shows that this attack is potent and widespread: dozens of organizations,
ranging from ones with fewer than 100 employees to ones with over 1,000 employees,
experience lateral phishing attacks within the span of several months;
in total, \percentRandomOrgsWithPhish of a set of randomly sampled organizations experienced at least one lateral phishing incident within a seven-month timespan.
Furthermore, we estimate that over \percentSuccessfulATO of attackers successfully compromise at least one additional employee.
Even though our ground truth sources and detector face limitations that restrict their ability to uncover stealthy or narrowly targeted attacks,
our results nonetheless illuminate a prominent threat that currently affects many real-world organizations.

Examining the behavior of lateral phishers,
we explore and quantify the popularity of four recipient (victim) selection strategies.
Although our dataset's attackers target dozens to hundreds of recipients,
these recipients often include a subset of users with some relationship to the hijacked account (\eg fellow employees or recent contacts).
Additionally, we develop a categorization for the different levels of content tailoring displayed by our dataset's phishing messages.
Our categorization shows that while \percentIncidentsTargetedMsg of attacks deploy targeted messages,
most attacks opt for generic content that a phisher could easily reuse across multiple organizations.
In particular, we observe that lateral phishers rely predominantly on two common lures:
a pretext of a shared document and a fake warning message about a problem with the recipient's account.
Despite the popularity of non-targeted content,
nearly one-third of our dataset's attackers invest additional time and effort
to make their attacks more convincing and/or to evade detection;
and, over 80\% of attacks occur during the normal working hours of the hijacked account.

Ultimately, this work yields two contributions that expand our understanding of enterprise phishing and potential defenses against it.
First, we present a novel
detector that achieves an order-of-magnitude better performance than prior work,
while operating on a minimal data requirement (only leveraging historical emails).
Second, through the first large-scale characterization of lateral phishing,
we uncover the scale and success of this emerging class of attacks and shed light on
common strategies that lateral phishers employ.
Our analysis illuminates a prevalent class of enterprise attackers whose behavior does not fully match
the tactics of targeted nation-state attacks or industrial espionage.
Nonetheless, these lateral phishers still achieve success in the absence of new defenses,
and many of our dataset's attackers do exhibit some signs of sophistication and focused effort.

%% file: background.tex
\section{Background}\label{sec:background}
In a \emph{lateral phishing} attack, attackers use a compromised, but legitimate, email account
to send a phishing email to their victim(s).
The attacker's goals and choice of malicious payload can take a number of different forms,
from a malware-infected attachment, to a phishing URL, to a fake payment request.
Our work focuses on lateral phishing attacks that employ a malicious URL embedded in the email,
which is the most common exploit method identified in our dataset.

Listing~\ref{lst:example2}
shows an anonymized example of a lateral phishing attack from our study.
In this attack, the phisher tried to lure the recipient into clicking on a link under the false pretense of a new contract.
Additionally, the attacker also tried to make the deception more credible by responding to recipients who inquired about the email's authenticity;
and they also actively hid their presence in the compromised user's mailbox by deleting all traces of their phishing email.

Lateral phishing represents a dangerous
but understudied attack at the intersection of phishing and account hijacking.
Phishing attacks, broadly construed, involve an attacker crafting a deceptive email from any account (compromised or spoofed) to trick their victim into performing some action.
Account hijacking, also known as account takeover (ATO) in industry parlance,
involves the use of a compromised account for any kind of malicious means (\eg including spam).
While prior work primarily examines each of these attacks at a smaller scale and with respect to personal accounts,
our work studies the intersection of both of these
at a large scale and from the perspective of enterprise organizations.
In doing so, we expand our understanding of important enterprise threats,
avenues for defending against them,
and strategies used by the attackers who perpetrate them.

\lstset{frame = single,columns=fullflexible, xleftmargin=.02\columnwidth, xrightmargin=.02\columnwidth}

\begin{lstlisting}[caption={An anonymized example of a lateral phishing message that uses the lure of a fake contract document.},
label={lst:example2},linewidth=\columnwidth,breaklines=true,float,aboveskip=0pt,belowskip=-0.8\baselineskip]
From: "Alice" <alice@company.com>
To: "Bob" <bob@company.com>
Subject: Company X (New Contract)

New Contract

View Document [this text linked to a phishing website]

Regards,
Alice [signature]
\end{lstlisting}

\subsection{Related Work}
\label{sec:related}

\paragraph{Detection}
An extensive body of prior literature proposes numerous techniques for
detecting traditional phishing
attacks~\cite{fette2007learning, bergholz2008improved, abu2007comparison, garera2007framework, whittaker2010large},
as well as more sophisticated
spearphishing attacks~\cite{stringhini2015ain, duman2016emailprofiler, zhao2016optimizing, khonji2011mitigation, barracudaBEC}.
Hu et al.\ studied how to use social graph metrics to detect malicious emails sent from compromised
accounts~\cite{hu2016detecting}.
Their approach detects hijacked accounts with false positive rates between 20--40\%.
Unfortunately, in practice, many organizations handle tens of thousands of employee-sent emails per day,
so a false positive rate of 20\% would lead to thousands of false alerts each day.
IdentityMailer, proposed by Stringhini et al.~\cite{stringhini2015ain},
detects lateral phishing attacks by training behavior models based on timing patterns, metadata, and stylometry
for each user.
If a new email deviates from an employee's behavioral model, their system flags it as an attack.
While promising, their approach produces false positive rates in the range of 1--10\%,
which is untenable in practice given the high volume of benign emails and low base rate of phishing.
Additionally, their system requires training a behavioral model for each employee,
incurring expensive technical debt to operate at scale.

Ho et al.\ developed methods for detecting
lateral spearphishing by applying a novel anomaly detection algorithm
on a set of features derived from historical user login data and enterprise network traffic logs~\cite{ho2017detecting}.
Their approach detects both known and newly discovered attacks, with a false positive rate of 0.004\%.
However, organizations with less technical expertise often lack the infrastructure to comprehensively capture
the enterprise's network traffic, which this prior approach requires.
This technical prerequisite begs the question, can we practically detect lateral phishing attacks with
a more minimalist dataset: only the enterprise's historical emails?
Furthermore, their dataset reflects a single enterprise
that experienced only two lateral phishing attacks across a 3.5-year timespan,
which prevents them from characterizing the nature of lateral phishing at a general scale.

\paragraph{Characterization}
While prior work shows that attackers frequently use phishing to compromise accounts,
and that attackers occasionally conduct (lateral) phishing from these hijacked accounts,
few efforts have studied the nature of lateral phishing in depth and at scale.
Examining a sample of phishing emails, webpages, and compromised accounts from Google data sources,
one prior study of account hijacking
discovered that attackers often use these accounts
to send phishing emails to the account's contacts~\cite{bursztein2014handcrafted}.
However, they concluded that automatically detecting such attacks proves challenging.
Onaolapo et al.\ studied what attackers do with hijacked
accounts~\cite{onaolapo2016happens}, but they did not examine lateral phishing.
Separate from email accounts,
a study of compromised Twitter accounts found that infections appear
to spread laterally through the social network.
However their dataset did not allow direct observation of the lateral attack vector
itself~\cite{thomas2014twitter}, nor did it provide insights into the domain of compromised enterprise accounts (given the nature of social media).

\paragraph{Open Questions and Challenges}
Prior work makes clear that account compromise poses a significant and widespread problem.
This literature also presents promising defenses for enterprises that have sophisticated monitoring in place.
Yet despite these advances, several key questions remain unresolved.
Do organizations without comprehensive monitoring and technical expertise have a practical way to defend against lateral phishing attacks?
What common strategies and tradecraft do lateral phishers employ?
How are lateral phishers capitalizing on their control of legitimate accounts,
and what does their tactical sophistication say about the state of enterprise phishing?
This paper takes a step towards answering these open questions by
presenting a new detection strategy and a large-scale characterization of lateral phishing attacks.

\subsection{Ethics}\label{sec:ethics}

In this work, our team, consisting of researchers from academia and a large security company,
developed detection techniques using a dataset of historical emails and reported incidents from \totalOrgs organizations who are active customers of \cudaFullName.
These organizations granted \cuda permission to access their Office 365 employee mailboxes for the purpose of researching and developing defenses against lateral phishing.
Per \cuda's policies, all fetched emails are stored encrypted, and customers have the option of revoking access to their data at any time.

Due to the sensitivity of the data, only authorized employees at \cuda were allowed to access the data (under standard, strict access control policies).
No personally identifying information or sensitive data was shared with any non-employee of \cuda.
Our project also received legal approval from \cuda,
who had permission from their customers to analyze and operate on the data.

Once \cuda deployed a set of lateral phishing detectors
to production, any detected attacks were reported to customers in real time to prevent financial loss and harm.

%% file: dataset.tex
\newcommand{\numOrgsLeqOneHundred}{25\xspace}
\newcommand{\numOrgsLeqOneThousandGreaterOneHundred}{34\xspace}
\newcommand{\numOrgsGreaterOneThousand}{33\xspace}
\section{Data}\label{sec:dataset}
\begin{figure}[t]
\includegraphics[width=1.0\columnwidth]{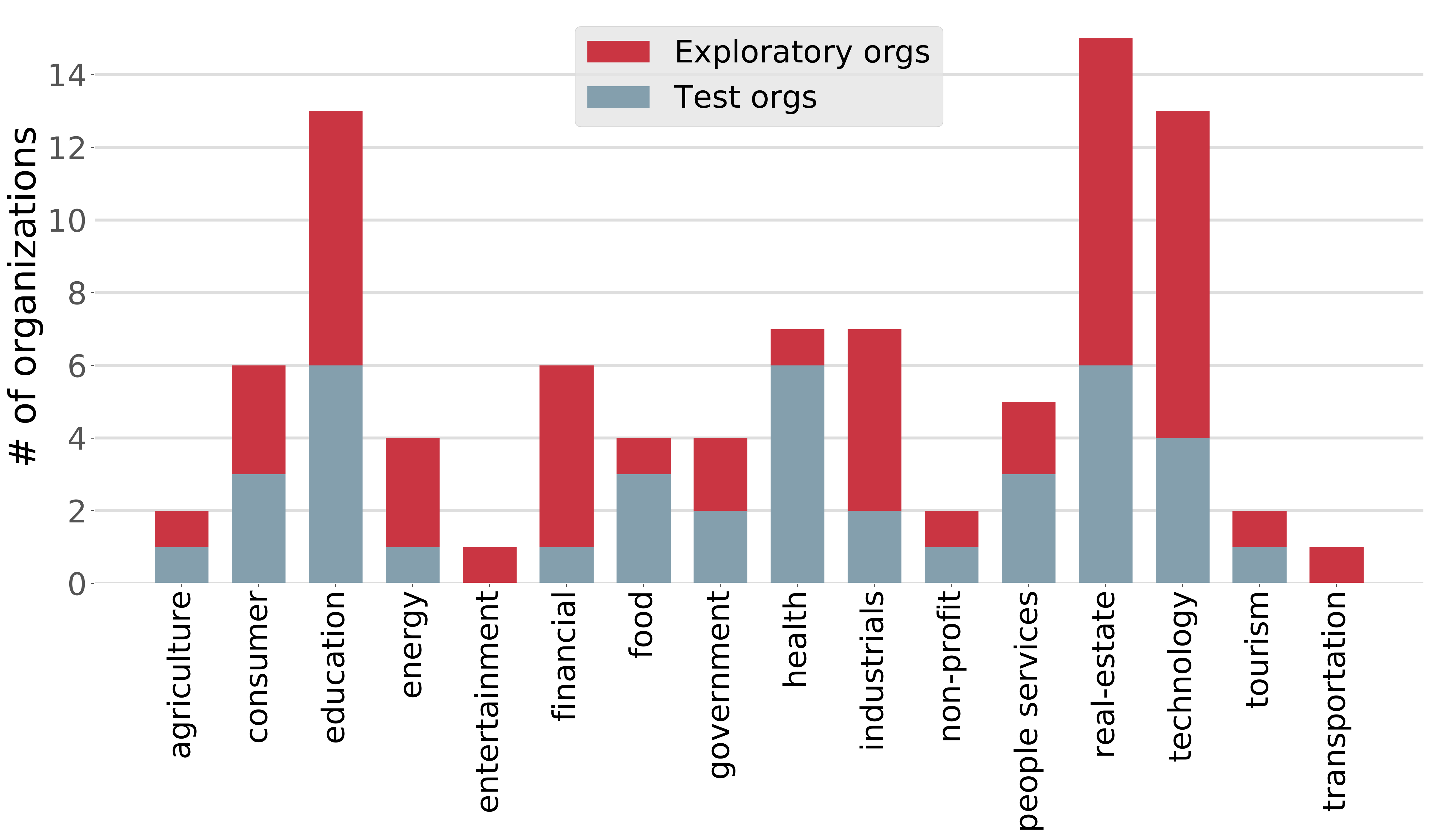}
\setlength{\abovecaptionskip}{-11pt}
\setlength{\belowcaptionskip}{-14pt}
\caption{Breakdown of the economic sectors across our dataset's \numTrainingOrgs \trainingOrgs
versus the \numTestOrgs \testOrgs.
}
\label{fig:orgSectorsByExploratoryAndTest}
\vspace*{0.1in}
\end{figure}

\begin{figure}[t]
\includegraphics[width=1.0\columnwidth]{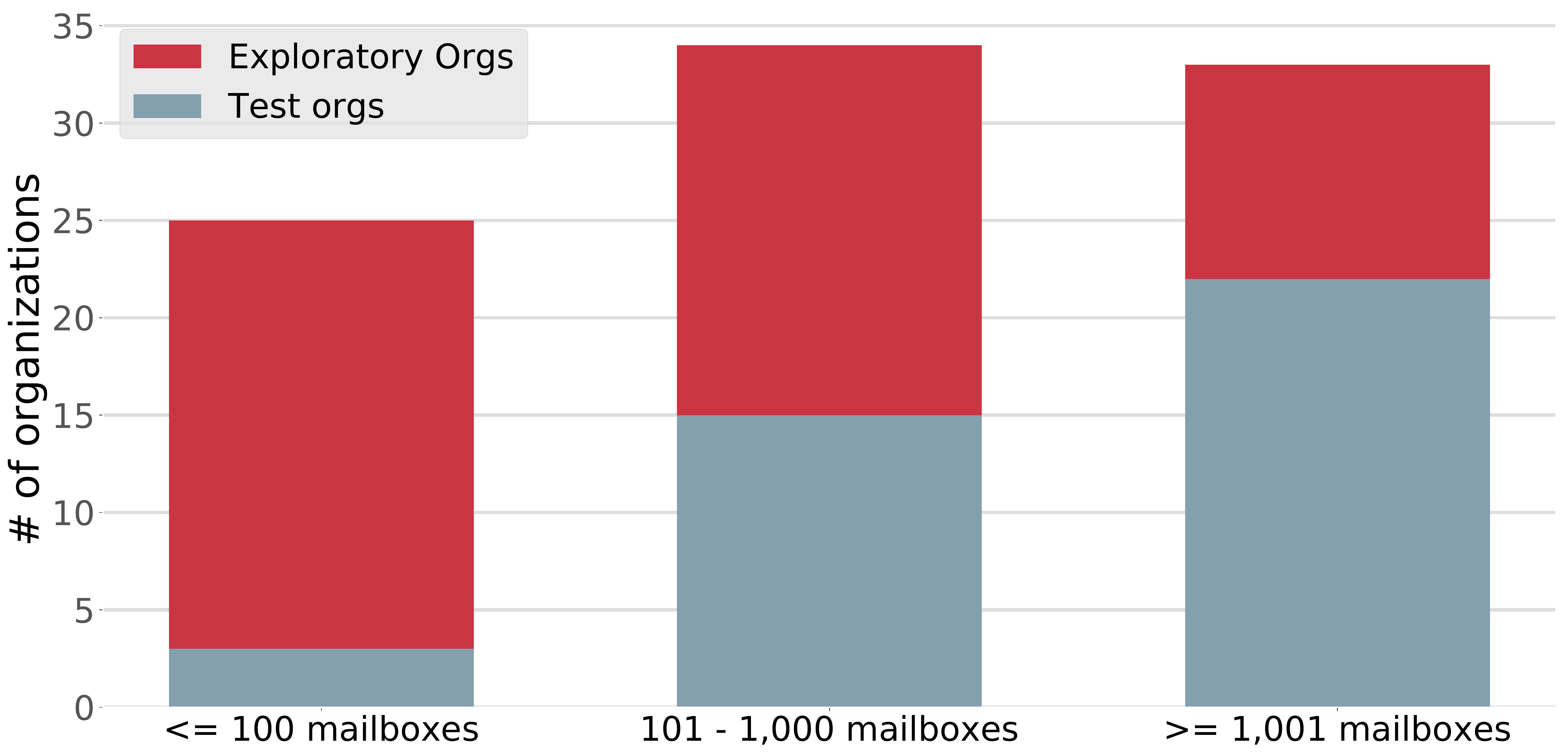}
\setlength{\abovecaptionskip}{-11pt}
\setlength{\belowcaptionskip}{-14pt}
\caption{Breakdown of the organization sizes across our dataset's \numTrainingOrgs \trainingOrgs
versus the \numTestOrgs \testOrgs.}
\label{fig:orgSizesByExploratoryAndTest}
\vspace*{0.1in}
\end{figure}

Our dataset consists of employee-sent emails from \totalOrgs English-language organizations;
\numPhishOrgsAPrioriWeKnew organizations came from randomly sampling enterprises that had reports of lateral phishing,
and \numRandomOrgs were randomly sampled from all organizations.
Across these enterprises,
\numOrgsLeqOneHundred organizations have 100 or fewer user accounts,
\numOrgsLeqOneThousandGreaterOneHundred have between 101--1000 accounts,
and \numOrgsGreaterOneThousand have over 1000 accounts.
Real-estate, technology, and education constitute the three most common industries
in our dataset, with 15, 13, and 13 enterprises respectively;
Figures~\ref{fig:orgSectorsByExploratoryAndTest} and~\ref{fig:orgSizesByExploratoryAndTest}
show the distribution of the economic sectors and sizes of our dataset's organizations,
broken down by \emph{\trainingOrgs} versus \emph{\testOrgs} (\S~\ref{sec:datastats}).

\subsection{Schema}\label{sec:schema}
The organizations in our dataset use Office 365 as their email provider.
At a high level, each email object contains: a unique Office 365 identifier;
the email's metadata (SMTP header information), which describes properties such as the email's
sent timestamp, recipients, purported sender, and subject;
and the email's \emph{body}, the contents of the email message in full HTML formatting.
Office 365's documentation describes the full schema of each email object~\cite{o365schema}.
Additionally, for each organization, we have a set of \emph{verified domains}:
domains which the organization has declared that it owns.

\subsection{Dataset Size}\label{sec:datastats}
Our dataset consists of \numEmails unique, employee-sent emails. 
To ensure our detection techniques generalized (Section~\ref{sec:evalmethodology}),
we split our data into a training dataset of emails from \numTrainingOrgs `\trainingOrgs' during \trainingMonths,
and a test dataset covering \evalMonths from \totalOrgs organizations.
Our test dataset consists of emails from the \numTrainingOrgs \trainingOrgs (but from a later, disjoint time period than our training dataset),
plus data from an additional, held-out set of \numTestOrgs `\testOrgs'.
We selected the \numTestOrgs \testOrgs via a random sample that we performed prior to analyzing any data.
Our training dataset has \numEmailsTrainDataset emails,
and our test dataset has \numEmailsTestDataset emails.
Both sets of organizations cover a diverse range of industries and sizes as shown in Figures~\ref{fig:orgSectorsByExploratoryAndTest} and~\ref{fig:orgSizesByExploratoryAndTest}.
The \trainingOrgs span a total of \numMailboxesTrainingOrgs user mailboxes that sent or received email,
and the \testOrgs have \numMailboxesTestOrgs mailboxes (based on the data from October 2018).\footnote{
The number of mailboxes is an upper bound
on the number of employees due to the use of mailing lists and aliases.
}

\subsection{Ground truth}\label{sec:groundtruth}
Our set of lateral phishing emails comes from two sources:
(1) attack emails reported to \cuda by an organization's security administrators, as well as attacks reported by users to their organization or directly to \cuda, and
(2) emails flagged by our detector
(\S\ref{sec:design}),
which we manually reviewed and labeled before including.

At a high-level, to manually label an email as phishing, or not,
we examined its message content, Office 365 metadata, and Internet Message Headers~\cite{rfcsmtpheaders}
to determine whether the email contained phishing content,
and whether the email came from a compromised account (versus an external account, which we do not treat as lateral phishing).
For example, if the Office 365 metadata showed that a copy of the email resided in the employee's \emph{Sent Items} folder,
or if its headers showed that the email passed the corresponding
SPF or DKIM~\cite{dkim} checks, then we considered the email to be lateral phishing.
Appendix~\S\ref{sec:manualLabeling} describes our labeling procedure in detail.

\begin{figure}[t]
\vspace{-.3cm}
\includegraphics[width=1.0\columnwidth]{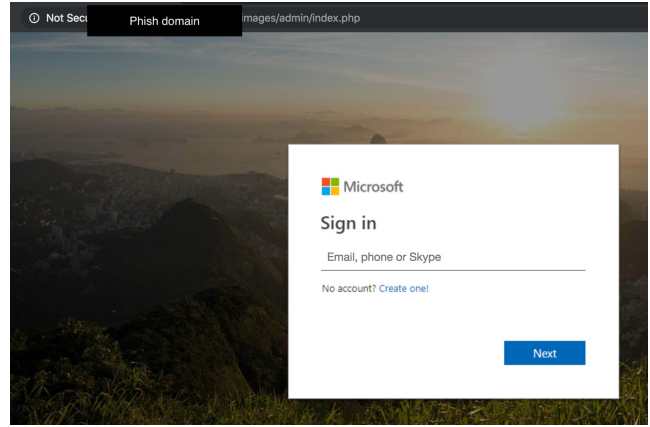}
\setlength{\abovecaptionskip}{-16pt}
\setlength{\belowcaptionskip}{-10pt}
\caption{An anonymized screenshot of the web page that a phishing URL in a lateral phishing email led to.}
\label{fig:eyeCandy}
\end{figure}

Additionally, for a small sample of URLs in these lateral phishing emails,
employees at \cuda accessed the phishing URL in a VM-contained browser to better understand the end goals of the attack.
To minimize potential harm and side effects,
these employees only visited phishing URLs which contained no unique identifiers
(\ie no random strings or user/organization information in the URL path).
To handle any phishing URLs that resided on URL-shortening domains,
we used one of \cuda's URL-expansion APIs that their production services already apply to email URLs,
and only visited suspected phishing links that expanded to a non-side-effect URL.
Most phishing URLs we explored led to a SafeBrowsing interstitial webpage,
likely reflecting our use of historical emails,
rather than what users would have encountered contemporaneously.
However, more recent malicious URLs consistently led to credential phishing websites
designed to look like a legitimate Office 365 login page
(the email service provider used by our study's organizations);
Figure~\ref{fig:eyeCandy} shows an anonymized example of one phishing website.

In total, our dataset contains \numPhishEmails lateral phishing emails (unique by subject, sender, and sent-time),
sent by \numHijackedAccounts hijacked employee accounts from \numPhishOrgs organizations.
\numReportedPhishEmails of these emails were reported by users, with the remainder found solely by our detector (\S~\ref{sec:design});
our detector also finds many of the user-reported attacks as well (\S~\ref{sec:evaluation}).
Among the user-reported attacks, \numAttachPhishEmails emails (from \numAttachATO hijacked accounts) contained a fake wire transfer or malicious attachment,
while the remaining \numURLPhishEmails emails used a malicious URL.

We focus our detection strategy on URL-based phishing,
given the prevalence of this attack vector.
This focus means that our analysis and detection techniques do not reflect the full space of lateral phishing attacks.
Despite this limitation, our dataset's attacks span dozens of organizations,
enabling us to study a prevalent class of enterprise phishing that poses an important threat in its own right.

%% file: design.tex
\section{Detecting Lateral Phishing}\label{sec:design}

Adopting the \emph{lateral attacker} threat model defined  by Ho et al.~\cite{ho2017detecting},
we focus on phishing emails sent by a compromised employee account,
where the attack embeds a malicious URL as the exploit (\eg leading the user to a phishing webpage).

We explored three strategies for detecting lateral phishing attacks,
but ultimately found that one of the strategies detected nearly all of the attacks identified by all three approaches.
At a high level, the two less fruitful strategies detected attacks by looking for emails that contained
(1) a rare URL and (2) a message whose text seemed likely to be used for phishing (\eg similar text to a known phishing attack).
Because our primary detection strategy detected all-but-\numPhishIncidentsFoundOnlyByTextDetectors of the attacks found by the other strategies,
while finding over ten times as many attacks,
we defer discussion of the two less successful approaches to Appendix~\ref{sec:textSimilarityDetection};
below, we focus on exploring the more effective strategy in detail.
In our evaluation,
we include the \numPhishIncidentsFoundOnlyByTextDetectors additional attacks found by the alternative approaches as false negatives for our detector.

\newcommand{\globalScoreUnrankedDomain}{10 million\xspace}
\newcommand{\largeRecipSize}{25\xspace}
\newcommand{\percentLargeRecipReportedTrainingATO}{95\%\xspace}

\paragraph{Overview}
We examined the user-reported lateral phishing incidents in our training dataset (\trainingMonths)
to identify widespread themes and behaviors that we could leverage in our detector.
Grouping this set of attacks by the hijacked account (\emph{ATO}) that sent them,
we found that \percentLargeRecipReportedTrainingATO of these ATOs sent phishing emails to
\largeRecipSize or more distinct recipients.\footnote{
To assess the generalizability of our approach,
our evaluation uses a withheld dataset, from a later timeframe and with new organizations (\S~\ref{sec:evaluation}).
}
This prevalent behavior, along with additional feature ideas inspired by the lure-exploit detection framework~\cite{ho2017detecting},
provide the basis for our detection strategy.
In the remainder of this section, we describe the features our detector uses,
the intuition behind these features,
and our detector's machine learning procedure for classifying emails.

Our techniques provide neither an all-encompassing approach to finding every attack,
nor guaranteed robustness against motivated adversaries trying to evade detection.
However, we show in Section~\ref{sec:evaluation} that
our approach finds hundreds of lateral phishing emails across dozens of real-world organizations,
while incurring a low volume of false positives.

\paragraph{Features}
Our detector extracts three sets of features.
The first set consists of two features that target the popular behavior we observed earlier: contacting many recipients.
Given an email, we first extract the number of unique recipients across the email's \texttt{To}, \texttt{CC}, and \texttt{BCC} headers.
Additionally, we compute the Jaccard similarity of this email's recipient set to
the closest set of historical recipients seen in any employee-sent email from the preceding month.
We refer to this latter (similarity) feature as the email's \featureRecipSimilarity score.

The next two sets of features draw upon the lure-exploit phishing framework proposed by Ho et al.~\cite{ho2017detecting}.
This framework posits that phishing emails contain two necessary components:
a `lure', which convinces the victim to believe the phishing email and perform some action;
and an `exploit': the malicious action the victim should execute.
Their work finds that using features that target both of these two components significantly improves a detector's performance.

To characterize whether a new email contains a potential phishing lure,
our detector extracts a single, lightweight boolean feature based on the email's text.
Specifically, \cuda provided us with a set of roughly 150 keywords and phrases that frequently occur in phishing attacks.
They developed this set of `phishy' keywords by extracting the link text from several hundred real-world phishing emails
(both external and lateral phishing) and selecting the (normalized) text that occurred most frequently among these attacks.
Thematically, these suspicious keywords convey a call to action that entices the recipient to click a link.
For our `lure' feature, we extract a boolean value that indicates whether an email contains any of these phishy keywords.

Finally, we complete our detector's feature set by extracting two features that capture whether an email might contain an exploit.
Since our work focuses on URL-based attacks, this set of features reflects whether the email contains a potentially dangerous URL.

First, for each email, we extract a \featureglobalrep feature that quantifies the rarest URL an email contains.
Given an email, we extract all URLs from the email's body and ignore URLs if they fall under two categories:
we exclude all URLs whose domain is listed on the organization's \emph{verified domain} list (\S~\ref{sec:schema}),
and we also exclude all URLs whose displayed, hyperlinked text exactly matches the URL of the hyperlink's underlying destination.
For example, in Listing~\ref{lst:example2}'s attack,
the displayed text of the phishing hyperlink was ``Click Here'',
which does not match the hyperlink's destination (the phishing site),
so our procedure would keep this URL.
In contrast, Alice's signature from Listing~\ref{lst:example2} might contain a link to her personal website,
\eg\ \texttt{www.alice.com};
our procedure would ignore this URL, since the displayed text of \texttt{www.alice.com} matches the hyperlink's destination.

This latter filtering criteria makes the assumption that a phishing URL will attempt to obfuscate itself,
and will not display the true underlying destination directly to the user.
After these filtering steps, we extract a numerical feature by mapping each remaining URL to its registered domain,
and then looking up each domain's ranking on the Cisco Umbrella Top 1 Million sites~\cite{ciscoUmbrella};\footnote{
We use a list fetched in early March 2018 for our feature extraction, but
in practice, one could use a continuously updated list.}
for any unlisted domain, we assign it a default ranking of \globalScoreUnrankedDomain.  
We treat two special cases differently.
For URLs on shortener domains, our detector attempts to recursively resolve the shortlink to its final destination.
If this resolution succeeds, we use the global ranking of the final URL's domain;
otherwise, we treat the URL as coming from an unranked domain (\globalScoreUnrankedDomain).
For URLs on content hosting sites (\eg Google Drive or Sharepoint),
we have no good way to determine its suspiciousness without fetching the content and analyzing it
(an action that has several practical hurdles).
As a result, we treat all URLs on content hosting sites as if they reside on unranked domains.

After ranking each URL's domain, we set the email's \featureglobalrep feature to be the worst (highest) domain ranking among its URLs.
Intuitively, we expect that phishers will rarely host phishing pages on popular sites,
so a higher \featureglobalrep indicates a more suspicious email.
In principle a motivated adversary could evade this feature;
\eg if an adversary can compromise one of the organization's verified domains,
they can host their phishing URL from this compromised site and avoid an accurate ranking.
However, we found no such instances in our set of user-reported lateral phishing.
Additionally, since the goal of this paper is to begin exploring practical detection techniques, and
develop a large set of lateral phishing incidents for our analysis, this feature suffices for our needs.

In addition to this global reputation metric,
we extract a local metric that characterizes the rareness of a URL with respect to the domains of URLs that an organization's employees typically send.
Given a set of URLs embedded within an email,
we map each URL to its fully-qualified domain name (FQDN)
and count the number of days from the preceding month where at least one employee-sent email included a URL on the FQDN.
We then take the minimum value across all of an email's URLs; we call this minimum value the \featurelocalrep feature.
Intuitively, suspicious URLs will have both a low global reputation and a low local reputation.
However, our evaluation (\S~\ref{sec:evalviral}) finds that this \featurelocalrep feature adds little value:
URLs with a low \featurelocalrep value almost always have a low \featureglobalrep value,
and vice versa.

\paragraph{Classification}
To label an email as phishing or not,
we trained a Random Forest classifier~\cite{randomforest} with the aforementioned features.
To train our classifier, we take all user-reported lateral phishing emails in our training dataset,
and combine them with a set of likely-benign emails.
We generate this set of ``benign'' emails by randomly sampling a subset of the training window's emails that have not been reported as phishing;
we sample 200 of these benign emails for each attack email to form our set of benign emails for training.
Following standard machine learning practices,
we selected both the hyperparameters for our classifier and the exact downsampling ratio (200:1)
using cross-validation on this training data.
Appendix~\ref{sec:hyperparameters} describes our training procedure in more detail.

Once we have a trained classifier, given a new email,
our detector extracts its features, feeds the features into this classifier, and outputs the classifier's decision.

%% file: evaluation.tex
\newcommand{\recallTerm}{Detection Rate\xspace}

\newcommand{\numAlertIncidentsTrainViral}{198\xspace}
\newcommand{\numPhishIncidentsTrainViral}{62\xspace}
\newcommand{\numFPIncidentsTrainViral}{136\xspace}
\newcommand{\percentRecallTrainViral}{88.6\%\xspace}

\newcommand{\numDetectedKnownIncidentsTrain}{34\xspace}
\newcommand{\numDetectedNewIncidentsTrain}{28\xspace}
\newcommand{\numMissedKnownIncidentsTrain}{8\xspace}
\newcommand{\fpAggregateTrain}{136\xspace}
\newcommand{\fpPercentAggregateTrain}{0.00053\%\xspace}
\newcommand{\precisionAggregateTrain}{31.3\%\xspace}

\newcommand{\numAlertIncidentsTestViral}{408\xspace}
\newcommand{\numPhishIncidentsTestViral}{96\xspace}
\newcommand{\numNewPhishIncidentsTestViral}{49\xspace}
\newcommand{\denomPhishIncidentsTestViral}{110\xspace}
\newcommand{\numFPIncidentsTestViral}{312\xspace}
\newcommand{\percentRecallTestViral}{87.3\%\xspace}
\newcommand{\percentFPTestViral}{0.00035\%\xspace}

\newcommand{\numDetectedKnownIncidentsTest}{47\xspace}
\newcommand{\numDetectedNewIncidentsTest}{49\xspace}
\newcommand{\numMissedKnownIncidentsTest}{14\xspace}
\newcommand{\fpAggregateTest}{316\xspace}
\newcommand{\precisionAggregateTest}{23.3\%\xspace}


\newcommand{\totalMissedIncidents}{22\xspace}

\section{Evaluation}\label{sec:evaluation}

In this section we evaluate our lateral phishing detector.  We first
describe our testing methodology, and then show how well
the detector performs on millions of emails from over 90
organizations.  Overall, our detector has a high
detection rate, generates few false positives, and detects many new attacks.

\subsection{Methodology}\label{sec:evalmethodology}

\paragraph{Establishing Generalizability}
As described earlier in Section~\ref{sec:datastats},
we split our dataset into two disjoint segments:
a \emph{training dataset} consisting of emails
from the \numTrainingOrgs \trainingOrgs during \trainingMonths and a \emph{test dataset}
from \totalOrgs enterprises during \evalMonths;
in~\S~\ref{sec:evalviral}, we show that our detector's performance remains the same if our test dataset contains only the emails from the  \numTestOrgs withheld \testOrgs.
Given these two datasets, we first trained our classifier and tuned its hyperparameters via cross validation on our training dataset (Appendix~\ref{sec:hyperparameters}).
Next, to compute our evaluation results, we ran our detector on each month of the held-out test dataset.
To simulate a classifier in production,
we followed standard machine learning practices and used a continuous learning procedure to update our detector each month~\cite{retraining}.
Namely, at the end of each month,
we aggregated the user-reported and detector-discovered phishing emails from all previous months into a new set of phishing `training' data;
and, we aggregated our original set of randomly sampled benign emails with our detector's false positives from all previous months to form a new benign `training' dataset.
We then trained a new model on this aggregated training dataset and used this updated model to classify the subsequent month's data.
However, to ensure that any tuning or knowledge we derived from the training dataset did not bias or overfit our classifier,
we did not alter any of the model's hyperparameters or features during our evaluation on the test dataset.

Our evaluation's temporal-split between the training and test datasets,
along with the introduction of new data from randomly withheld organizations into the test dataset,
follows best practices that recommend this approach over a
randomized cross-validation evaluation~\cite{allix2015your, miller2016reviewer, pendlebury2018tesseract}.
A completely randomized evaluation (\eg cross-validation)
risks training on data from the future and testing on the past,
which might lead us to overestimate the detector's effectiveness.
In contrast, our methodology evaluates our detector
with fresh data from a ``future'' time period and introduces \numTestOrgs new organizations,
neither of which our detector saw during training time;
this also reflects how a detector operates in practice.

\paragraph{Alert Metric (Incidents)}
We have several choices for modeling our detector's alert generation process
(\ie how we count \emph{distinct} attacks).
For example, we could evaluate our detector's performance in terms of how many unique emails it correctly labels.
Or, we could measure our detector's performance in terms of how many distinct employee accounts it marks as compromised
(modeling a detector that generates one alert per account and suppresses the rest).
Ultimately, we select a notion commonly used in practice, that of an \emph{incident},
which corresponds to a unique (subject, sender email address) pair.
At this granularity, our detector's alert generation model produces a single alert per unique (subject, sender) pair.
This metric avoids biased evaluation numbers that overemphasize compromise incidents that generate many identical emails during a single attack.
For example, if there are two incidents, one which generates one hundred emails to one recipient each, and another which generates one email to 100 recipients,
a detector's performance on the hundred-email incident will dominate the result if we count attacks at the email level.

In total, our training dataset contains \numReportedTrainIncidents lateral phishing incidents from our user-reported ground truth sources, and
our test dataset contains \numReportedTestIncidents user-reported incidents.
Our detector finds an additional \numNewlyDiscoveredIncidents unreported incidents (row 2 of Table~\ref{table:evaluation}).

\begin{table}
\centering\small
\resizebox{\columnwidth}{!}{
\begin{tabular}{lrr}
& \multicolumn{1}{c}{\bf Training } & \multicolumn{1}{c}{\bf Testing } \\
{\bf Metric} & April -- June 2018 & July -- October 2018 \\
\toprule
Organizations & 52 Exploratory & 52 Exploratory \\
 & & + 40 Test \\
\midrule
Detected Known Attacks & \numDetectedKnownIncidentsTrain & \numDetectedKnownIncidentsTest \\
Detected New Attacks &  \numDetectedNewIncidentsTrain & \numDetectedNewIncidentsTest  \\
Missed Attacks (FN)  & \numMissedKnownIncidentsTrain  &  \numMissedKnownIncidentsTest \\
\recallTerm   & \percentRecallTrainViral & \recallAggregateTest   \\
\midrule
Total Emails & \numEmailsTrainDataset &  \numEmailsTestDataset \\
False Positives (FP)      &  \fpAggregateTrain &  \fpAggregateTest \\
False Positive Rate &  \fpPercentAggregateTrain &  \fpPercentAggregateTest \\
Precision   & \precisionAggregateTrain &  \precisionAggregateTest \\
\bottomrule
\end{tabular}
}
\caption{
Evaluation results of our detector.
`Detected Known Attacks' shows the number of incidents
that our detector identified, and were also
reported by an employee at an organization.
`Detected New Attacks' shows the number of incidents
that our detector identified,
but were not reported by anyone.
`Missed Attacks (FN)' shows all incidents either reported by a user or found by any of our detection strategies,
but our detector marked it as benign (false negative).
Of the \totalMissedIncidents incidents our detector misses,
\numAttachIncidents are attachment-based attacks, a threat model which our detector explicitly does not target
but which we include in our FN and \recallTerm results for completeness.
}
\label{table:evaluation}
\vspace{-.3cm}
\end{table}

\subsection{Detection Results}\label{sec:evalviral}
Table~\ref{table:evaluation} summarizes the performance metrics for our detector.
We use the term \emph{\recallTerm} to refer to the percentage of lateral phishing incidents that our detector finds,
divided by all known attack incidents in our dataset (\ie any user-reported incident and any incident found by any detection technique we tried).
For completeness, we include the \numAttachIncidents attachment-based incidents in our False Negative and \recallTerm computations,
which our detector obviously misses since we designed it to catch URL-based lateral phishing.
Additionally, we also include, as false negatives, \numTrainIncidentsTextFoundFN training incidents 
that our less successful detectors identified (Appendix~\ref{sec:textSimilarityDetection});
these two alternative strategies did not find any new attacks in the test dataset.
Thus, the \emph{\recallTerm} reflects a best-effort assessment that potentially overestimates the true positive rate of our detector,
since we have an imperfect ground truth that cannot account for narrowly targeted attacks that go unreported by users.
\emph{Precision} equals the percent of attack alerts (incidents) produced by our detector divided by the total number of alerts our detector generated
(attacks plus false positives).

\paragraph{Training and Tuning}
On the training dataset, our detector correctly identified
\numPhishIncidentsTrainViral out of \numTrainIncidents lateral phishing incidents (\percentRecallTrainViral),
while generating a total of \numPhishIncidentsTrainViral false positives
(on 25.7 million employee-sent emails).

Our PySpark Random Forest classifier exposes a built-in estimate of each feature's relative importance~\cite{pysparkRandomForest},
where each feature receives a score between 0.0--1.0 and the sum of all the scores adds up to 1.0.
Based on these feature weights,
our model places the most emphasis on the \featureglobalrep feature, giving it a weight of 0.42,
and the email's `number of recipients' feature (0.34).
In contrast, our model essentially ignores our \featurelocalrep,
assigning it a score of 0.01,
likely because most globally rare domains tend to also be locally rare.
Of the remaining features, the \featureRecipSimilarity feature has a weight of 0.17
and the `phishy' keyword feature has a weight of 0.06.

\newcommand{\numOrgsLEQTenFP}{82\xspace}
\newcommand{\numOrgsNoFP}{44\xspace}

\newcommand{\testRecallOnlyTestOrgs}{91.0\%\xspace}
\newcommand{\testPrecisionOnlyTestOrgs}{23.1\%\xspace}
\newcommand{\testFPROnlyTestOrgs}{0.00038\%\xspace}    

\paragraph{Test Dataset}
Our detector correctly identified \numPhishIncidentsTestViral lateral phishing incidents
out of the \denomPhishIncidentsTestViral test incidents (\percentRecallTestViral) across our ground truth dataset.
Additionally, our detector discovered \numNewPhishIncidentsTestViral incidents
that, according to our ground truth, were not reported by a user as phishing.
With respect to its cost, our detector generated \numFPIncidentsTestViral total false positives
across the entire test dataset (a false positive rate of less than~\percentFPTestViral,
assuming that emails not identified as an attack by our ground truth are benign).
Across our test dataset, \numOrgsLEQTenFP out of the \totalOrgs organizations
accumulated 10 or fewer false positives across the entire four month window,
with \numOrgsNoFP organizations encountering zero false positives across this timespan.
In contrast, only three organizations had more than 40 total false positives across all four months
(encountering 44, 66, and 83 false positives, respectively).
Our detector achieves similar results if we evaluate on just the data from our \numTestOrgs withheld \testOrgs,
with a \recallTerm of \testRecallOnlyTestOrgs, a precision of \testPrecisionOnlyTestOrgs,
and a false positive rate of \testFPROnlyTestOrgs.

\paragraph{Bias and Evasion}
We base our evaluation numbers on the best ground truth we have:
a combination of all user-reported lateral phishing incidents (including some attacks outside our threat model),
and all incidents discovered by any detection technique we tried
(which includes two approaches orthogonal to our detector's strategy). 
This ground truth suffers from a bias towards phishing emails that contact many potential victims,
and attacks that users can more easily recognize.
Additionally, since our detector focuses on URL-based exploits,
our dataset of attacks likely underestimates the prevalence of non-URL-based phishing attacks,
which come solely from user-reported instances in our dataset.
As a result, our work does not capture the full space of lateral phishing attacks,
such as ones where the attacker targets a narrow, select set of victims with stealthily deceptive content.
Rather, given that our detector identifies many known and unreported attacks,
while generating only a few false positives per month,
we provide a starting point for practical detection that future work can extend.
Moreover, even if our detector does not capture every possible attack,
the fact that the attacks in our dataset span dozens of different organizations, across a multi-month timeframe,
allows us to illuminate a class of understudied attacks that many enterprises currently face.

Aside from obtaining more comprehensive ground truth,
more work is needed to explore defenses against potential evasion attacks.
Attackers could attempt to evade our detector by targeting different features we draw upon,
such as the composition or number of recipients they target.
Against many of these evasion attacks, future work could leverage additional features and data,
such as the actions a user takes within an email account
(\eg reconnaissance actions, such as unusual searches, that indicate an attacker mining the account for targeted recipients to attack)
or information from the user's account log-on
(\eg the detector proposed by Ho et al.\ used an account's login IP address~\cite{ho2017detecting} to detect lateral phishing).
At the same time, future work should study which evasion attacks remain economically feasible for attackers to conduct.
For example, an attacker could choose to only target a small number of users in the hopes of evading our detector;
but even if this evasion succeeded,
the conversion rate of fooling a recipient might be so low that the attack ultimately fails to compromise an economically viable number of victims.
Indeed, as we explore in the following section (\S~\ref{sec:characterization}),
the attackers captured in our dataset already engage in a range of different behaviors,
including a few forms of sophisticated, manual effort to increase the success of their attacks.

%% file: characterization.tex
\begin{table}[]
\centering
\begin{tabular}{@{}lr@{}}
\toprule
 \multicolumn{2}{c}{\bf Scale and Success} \\
\midrule
 \# distinct phishing emails & \numPhishEmails \\
 \# incidents & \numIncidents \\
 \# ATOs & \numHijackedAccounts \\
 \# organizations w/ 1+ incident & \numPhishOrgs \\
 \# phishing recipients & \numPhishRecipAll \\
 \% successful ATOs & \percentSuccessfulATO \\
\# employee recip (average) for compromise & \numEmployeeRecipForSuccessfulInfectee \\
\bottomrule
\end{tabular}
\setlength{\belowcaptionskip}{-8pt}
\caption{Summary of the scale and success of the lateral phishing attacks in our dataset (\S~\ref{sec:scale}).
}
\label{table:phishSummary}
\end{table}

\section{Characterizing Lateral Phishing}\label{sec:characterization}
In this section, we conduct an analysis of real-world lateral phishing using
all known attacks across our entire dataset (both training and test).
During the seven month timespan, a total of 33 organizations experienced lateral phishing attacks,
with the majority of these compromised organizations experiencing multiple incidents.
Examining the thematic message content and recipient targeting strategies
of the attacks,
our analysis suggests that most lateral phishers in our dataset
do not actively mine a hijacked account's emails to craft personalized spearphishing attacks.
Rather, these attackers operate in an opportunistic fashion and rely on commonplace phishing content.
This finding suggests that the space of enterprise phishing has expanded beyond its historical association with sophisticated APTs
and nation-state adversaries.

At the same time, these attacks nonetheless succeed, and a significant fraction of attackers do exhibit some signs of sophistication and attention to detail.
As an estimate of the success of lateral phishing attacks,
at least \percentSuccessfulATO of our dataset's attackers successfully compromise at least one other employee account.
In terms of more refined tactics, \percentSophisticatedATO of lateral phishers
invest some manual effort in evading detection or increasing their attack's success rate.
Additionally, over 80\% of the attacks in our dataset occur during the normal working hours of the hijacked account.
Taken together, our results suggest that lateral phishing attacks pose a prevalent enterprise threat that
still has room to grow in sophistication.

In addition to exploring attacks at the incident granularity (as done in \S~\ref{sec:evaluation}),
this section also explores attacks at the granularity of a lateral phisher (hijacked account)
when studying different attacker behaviors.
As described in Section~\ref{sec:background}, industry practitioners often refer to such hijacked accounts as ATOs,
and throughout this section,
we use the terms \emph{hijacked account}, \emph{lateral phisher}, and \emph{ATO} synonymously.


\subsection{Scale and Success of Lateral Phishing}\label{sec:scale}

\begin{figure}
\includegraphics[width=1.0\columnwidth]{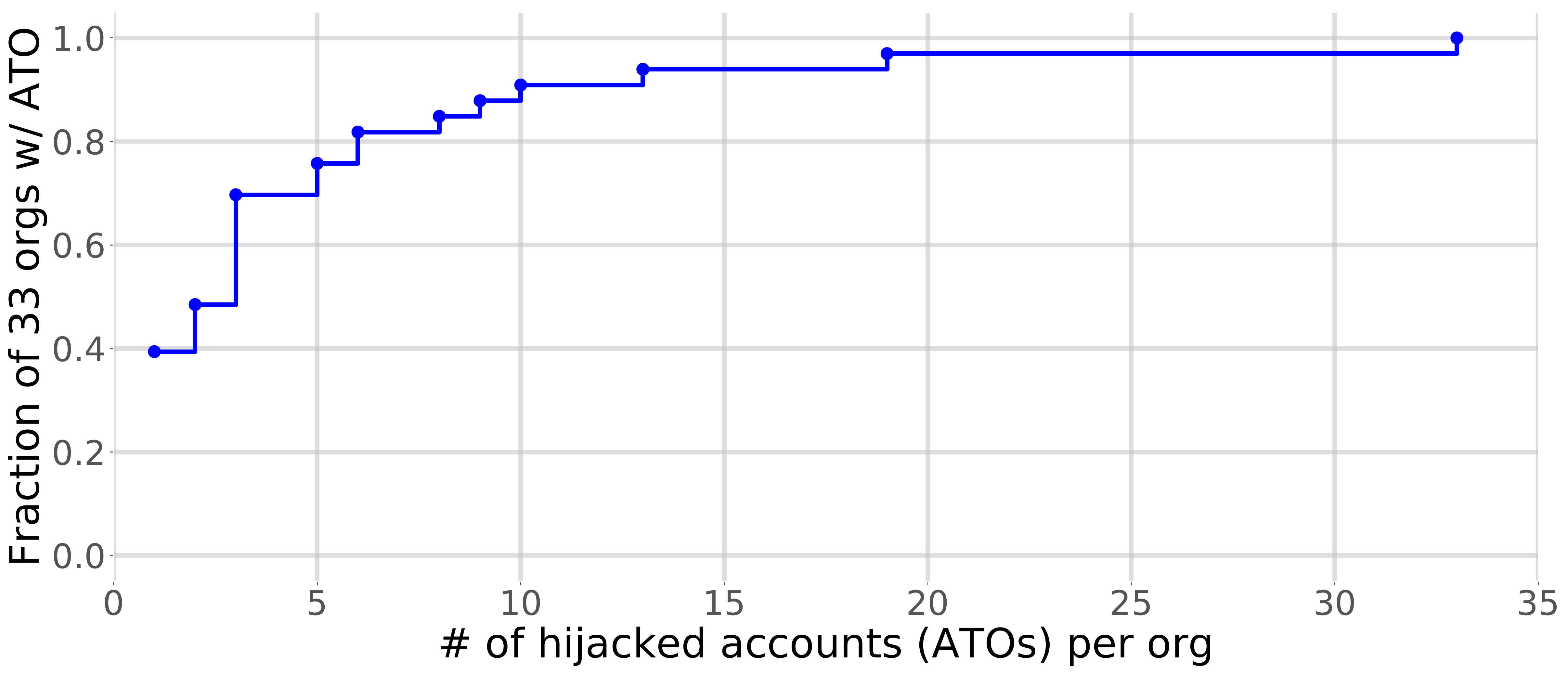}
\setlength{\abovecaptionskip}{-8pt}
\setlength{\belowcaptionskip}{-8pt}
\caption{Fraction of organizations with $x$ hijacked accounts that
sent at least one lateral phishing email.
\numOrgsWithSingleATO organizations had only 1 ATO;
the remaining \numOrgsWithMultiATO saw lateral phishing from 2+ ATOs
(\S~\ref{sec:scale}).
}
\label{fig:incidentsPerOrg}
\end{figure}

\paragraph{Scale}
Our dataset contains \numPhishEmails distinct lateral phishing emails
sent by \numHijackedAccounts hijacked accounts.\footnote{
Distinct emails are defined by having a fully unique tuple of
(sender, subject, timestamp, and recipients).
}
A total of \numPhishOrgs organizations in our dataset experience at least one lateral phishing incident:
\numPhishOrgsAPrioriWeKnew of these organizations came from sampling the set of enterprises with known lateral phishing incidents (\S~\ref{sec:dataset}),
while the remaining \numPhishOrgsFromSampling came from the \numRandomOrgs organizations we sampled from the general population.
Assuming our random sample reflects the broader population of enterprises,
over \percentRandomOrgsWithPhish of organizations experience at least one lateral phishing incident within a 7 month timespan.
Furthermore, based on Figure~\ref{fig:incidentsPerOrg},
over 60\% of the compromised organizations in our dataset experienced lateral phishing attacks from at least two hijacked employee accounts.
Given that our set of attacks likely contains false negatives (thus underestimating the prevalence of attacks),
these numbers illustrate that lateral phishing attacks are widespread across enterprise organizations.

\newcommand{\AlicePhish}{$P_{A}$\xspace}
\newcommand{\BobPhish}{$P_{B}$\xspace}
\newcommand{\BobReply}{$\mathit{Reply}_{B}$\xspace}
\paragraph{Successful Attacks}
Given our dataset, we do not definitively know whether an attack succeeded.
However, we conservatively (under)estimate the success rate of lateral phishing using the methodology below.
Based on this procedure, we estimate that at least \percentSuccessfulATO of lateral phishers successfully compromise at least one new enterprise account.

Let Alice and Bob represent two different ATOs at the same organization,
where \AlicePhish and \BobPhish represent one of Alice's and Bob's phishing emails respectively,
and \BobReply represents a reply from Bob to a lateral phishing email he received from Alice.
Intuitively, our methodology concludes that Alice successfully compromised Bob if (1) Bob received a phishing email from Alice,
(2) shortly after receiving Alice's phish, Bob then subsequently sent his own phish,
and (3) we have strong evidence that the two employees' phishing emails are related (reflected in criteria 3 and 4 below).

Formally, we say that \AlicePhish succeeded in compromising Bob's account if \emph{all} of the following conditions are true:
\begin{enumerate}
	\item Bob was a recipient of \AlicePhish
  \item After receiving \AlicePhish, Bob subsequently sent his own lateral phishing emails (\BobPhish)
	\item Either of the following two conditions are met:
    \begin{enumerate}
      \item \BobPhish and \AlicePhish used similar phishing content:
        if the two attacks used identical subjects or if both of the phishing URLs they used belonged to the same fully-qualified domain
      \item Bob sent a reply (\BobReply) to \AlicePhish, where his reply suggests he fell for Alice's attack and where Bob sent \BobReply prior to his own attack (\BobPhish)
    \end{enumerate}
	\item Either of the following two conditions are met:
    \begin{enumerate}
      \item \BobPhish was sent within two days after Bob received \AlicePhish
      \item \BobPhish and \AlicePhish used identical phishing messages or their phishing URLs' paths followed nearly identical structures
        (\eg `\texttt{http://X.com/z/office365/index.html}' vs.
        `\texttt{http://Y.com/z/office365/index.html}')
    \end{enumerate}
\end{enumerate}
Unpacking the final criteria (\#4), in the first case (4.a),
we settled on a two-day interarrival threshold based on prior literature~\cite{msrEmailInfoFlow, kanich2008spamalytics},
which suggests that 50\% of users respond to an email within 2 days and roughly 75\% of users who click on a spam email do so within 2 days.
Assuming that phishing follows similar time constants for how long it takes a recipient to take action,
2 days represented a conservative threshold to establish a link between \AlicePhish and \BobPhish.
At the same time, both prior works show there exists a long tail of users who take weeks to read and act on an email.
The second part (4.b) attempts to address this long tail by raising the similarity requirements between Alice and Bob's attacks
before concluding that former caused the latter.
For successful attackers labeled by heuristic~4.b, the longest observed
time gap between \AlicePhish and \BobPhish is 17 days,
which falls within a plausible timescale based on the aforementioned literature.

From this methodology, we conclude that \numSuccessfulATOs ATOs successfully compromised at least \numInfecteeATOs future ATOs.
While our procedure might erroneously identify cases where an attacker has concurrently compromised both Alice and Bob (rather than compromising Bob's account via Alice's),
the first two criteria
(requiring Bob to be a recent recipient of Alice's phishing email) help reduce this error.
Our procedure likely underestimates the general success rate of lateral phishing attacks,
since it does not identify successful attacks where the attacker does not subsequently use Bob's account to send phishing emails,
nor does it account for false negatives in our dataset or attacks outside of our visibility
(\eg compromise of recipients at external organizations).

\begin{figure*}[t]
\centering
\begin{minipage}[b]{0.48\linewidth}
\centering
\includegraphics[width=.9\textwidth]{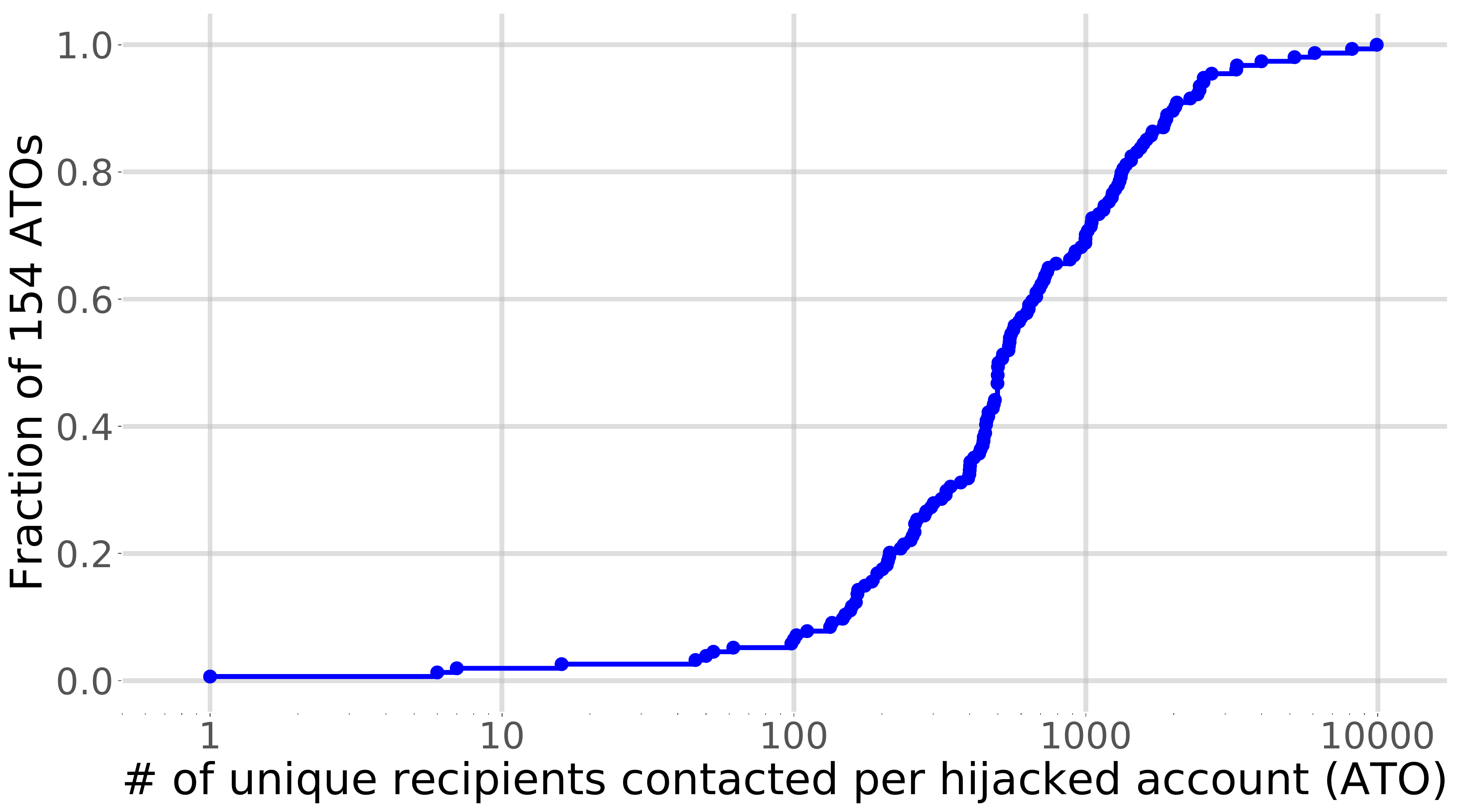}
\end{minipage}
\quad
\begin{minipage}[b]{0.48\linewidth}
\centering
\includegraphics[width=.9\textwidth]{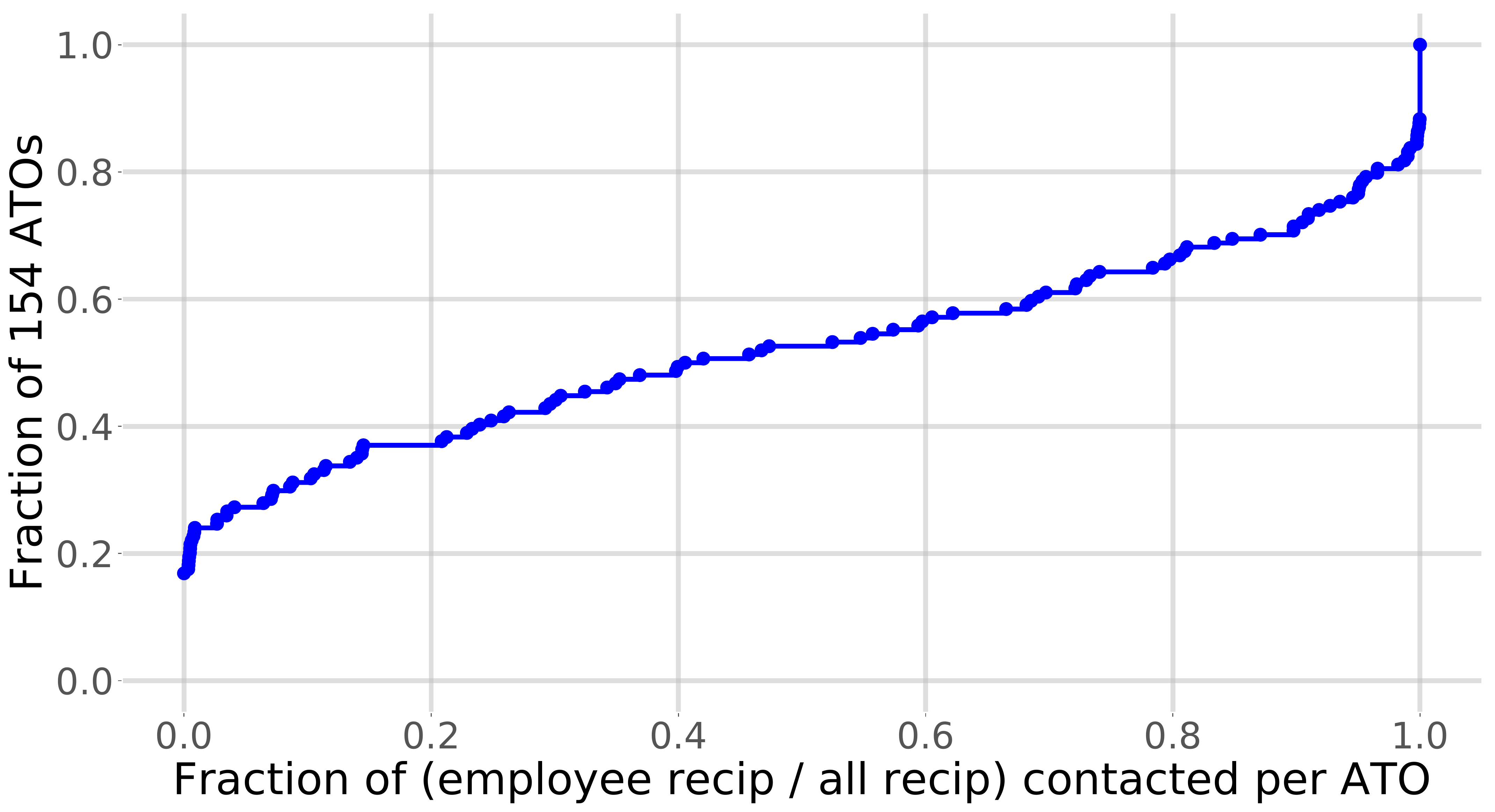}
\end{minipage}
\setlength{\belowcaptionskip}{-8pt}
\caption{The left CDF shows the distribution of the total number of phishing recipients per ATO.
The right CDF shows the fraction of ATOs where $x$\% of their total recipient set consists of fellow employees.
}
\label{fig:recipDistribution}
\vspace*{0.1in}
\end{figure*}


\subsection{Recipient Targeting}\label{sec:targeting}
\newcommand{\numPhishFromGeqFiftyRecipATO}{1,212\xspace}

In this section, we estimate the conversion rate of our dataset's lateral phishing attacks,
and discuss four recipient targeting strategies that reflect the behavior of most attackers in our dataset.

\paragraph{Recipient Volume and Estimated Conversation Rate}
Cumulatively, the lateral phishers in our dataset contact \numPhishRecipAll unique recipients,
where \numPhishRecipEmployee belong to the same organization as the ATO.
As shown in Figure~\ref{fig:recipDistribution},
more than \percentATOsTotalRecipLarge of the attackers send their phishing emails to over 100 recipients;
with respect to the general population of all lateral phishers,
this percentage likely overestimates the prevalence of high ``recipient-volume'' attackers,
since our detector draws on recipient-related features.

Targeting hundreds of people gives attackers a larger pool of potential victims,
but it also incurs a risk that a recipient will detect and flag the attack
either to their security team or their fellow recipients (\eg via Reply-All).
To isolate their victims and minimize the ability for fellow recipients to warn each other,
we found that attackers frequently
contact their recipients via a mass \texttt{BCC} or through many individual emails.
Aside from this containment strategy, we also estimate that our dataset's lateral phishing attacks have a difficult time fooling an individual employee,
and thus might require targeting many recipients to hijack a new account.
Earlier in Section~\ref{sec:scale},
we found that \numSuccessfulATOs ATOs successfully compromised \numInfecteeATOs new accounts.
Looking at the number of accounts they successfully hijacked divided by the number of fellow employees they targeted,
the median conversation rate for our attackers was
one newly hijacked account per
\numEmployeeRecipForSuccessfulInfectee fellow employees;
the attacker with the best conversation rate contacted an average of \minNumEmployeeRecipForSuccessfulInfectee employees per successful compromise.

We caution that our method for determining whether an attack succeeded (\S~\ref{sec:scale}) does not cover all cases,
so our conversation rate
might also underestimate the success of these attacks in practice.
But if our estimated conversion rate accurately approximates the true rate,
it would explain why these attackers contact so many recipients, despite the increased risk of detection.

\paragraph{Recipient Targeting Strategies}
Anecdotally, we know that some lateral phishers select their set of victims
by leveraging information in the hijacked account to target familiar users;
for example, sending their attack to a subset of the account's ``Contact Book''.
Unfortunately our dataset does not include
information about any reconnaissance actions that an attacker performed to select their phishing recipients
(\eg explicitly searching through a user's contact book or recent recipients).

Instead, we empirically explore the recipient sets across our dataset's attackers to identify plausible strategies for how these attackers might have chosen their set of victims.
Four recipient targeting strategies, summarized in Table~\ref{table:recipTargeting} (explained below),
reflect the behavior of all but six attackers in our dataset.
To help assess whether a recipient and the ATO share a meaningful relationship,
we compute each ATO's \emph{recent contacts}: the set of all email addresses
whom the ATO sent at least one email to in the 30 days preceding the ATO's phishing emails.
While some attackers (\percentATOTargetedRecipOverall) specifically target many of an account's recent contacts,
the majority of lateral phishers appear more interested in either contacting many arbitrary recipients
or sending phishing emails to a large fraction of the hijacked account's organization.

\begin{table}
\centering
\begin{tabular}{@{}lr@{}}
\toprule
 {\bf Recipient Targeting Strategy} & {\bf \# ATOs} \\ \midrule
 \atoRecipAgnostic & \numATOsRecipAgnostic \\
 \atoOrgWide & \numATOsOrgWide \\
 \atoLateralOrg & \numATOsLateralOrg \\
 \atoTargetedRecip & \numATOTargetedRecip \\
 Inconclusive & \numATOUnknownRecipTargeting \\
 \bottomrule
\end{tabular}
\setlength{\belowcaptionskip}{-8pt}
\caption{Summary of recipient targeting strategies per ATO (\S~\ref{sec:targeting}).
}
\label{table:recipTargeting}
\end{table}

\paragraph{\atoRecipAgnostic Attackers}
Starting with the least-targeted behavior,
\numATOsRecipAgnostic ATOs in our dataset sent their attacks to a wide range of recipients,
most of whom do not appear closely related to the hijacked account.
We call this group \emph{\atoRecipAgnostic attackers}, and identify them using two heuristics.

First, we categorize an attacker as \atoRecipAgnostic if less than 1\% of the recipients belong to the same organization as the ATO,
and further exploration of their recipients does not reveal a strong connection with the account.
Examining the right-hand graph in Figure~\ref{fig:recipDistribution},
\numATOsNoEmployeeRecip ATOs target recipient sets where less than 1\% of the recipients belong to the same organization as the ATO.
To rule out the possibility that these attackers' recipients are nonetheless related to the account,
we computed the fraction of recipients who appeared in each ATO's recent contacts;
for all of the \numATOsNoEmployeeRecip possible \atoRecipAgnostic ATOs,
less than \percentMaxContactBookOverlapRecipAgnostic of their attack's total recipients appeared in their recent contacts.
Among these \numATOsNoEmployeeRecip candidate \atoRecipAgnostic ATOs,
\numATOsNoEmployeeRecipAndManyTotalDomains of them contact recipients at 10 or more organizations
(unique recipient email domains),
\numATOsNoEmployeeRecipAndAllPersonalDomains of them exclusively target either Gmail or Hotmail accounts,
and the remaining \numATOsLateralOrg ATOs are best described as \atoLateralOrg attackers (below).\footnote{
Figure~\ref{fig:recipDomainsPerATO} in Appendix~\ref{sec:techreportfigures} shows the distribution of recipient domains contacted by all ATOs.
}
Excluding the \numATOsLateralOrg \atoLateralOrg attackers,
the \numATOsRecipAgnosticNoEmployeeCriteria ATOs identified by this first criteria
sent their attacks to predominantly external recipients,
belonging to either many different organizations or exclusively to personal email hosting services (\eg Gmail and Hotmail),
and only a small percentage of these recipients appeared in the ATO's recent contacts;
as such, we label these \numATOsRecipAgnosticNoEmployeeCriteria attackers as \atoRecipAgnostic.

Second, we expand our search for \atoRecipAgnostic attackers by searching for attackers
where less than 50\% of the ATO's total recipients also belong to the ATO's organization,
and where the ATO contacts recipients at many different organizations;
specifically, where the ATO's phishing recipients belonged to over twice as many unique domains
as all of the email addresses in ATO's recent contacts.
This search identified \numATOsMajorityExternalRecipAndManyTotalDomains ATOs.
To filter out attackers in this set who may have drawn on the hijacked account's recent contacts,
we exclude any ATO where over \percentMaxContactBookOverlapRecipAgnostic of their attack's total recipients also appeared in the ATO's recent contacts
(\percentMaxContactBookOverlapRecipAgnostic was the maximum percentage among ATOs from the first \atoRecipAgnostic heuristic).
After applying this last condition, our second heuristic identifies \numATOsRecipAgonisticSecondVersion \atoRecipAgnostic attackers.

Combining and deduplicating the ATOs from both criteria results in
a total of \numATOsRecipAgnostic \atoRecipAgnostic attackers (\percentATOsRecipAgnostic):
lateral phishers who predominantly target recipients without close relationships to the hijacked account or its organization.

\paragraph{\atoLateralOrg Attackers}
During our exploration of potential \atoRecipAgnostic ATOs,
we uncovered \numATOsLateralOrg attackers whom we label under a different category: \emph{\atoLateralOrg attackers}.
In both these cases, less than 1\% of the attacker's recipients belonged to the same organization as the ATO,
but each attacker's recipients did belong to organizations within the same industry as the ATO's organization.
This thematic characteristic among the recipients suggests a deliberate strategy to spread across organizations within the targeted industries,
so accordingly, we categorize them as \atoLateralOrg attackers.

\paragraph{\atoOrgWide Attackers}
Office 365 provides a ``Groups'' feature that
lists the different groups that an account belongs to~\cite{o365ContactsAPI}.
For some enterprises, this feature enumerates most, if not all, employees at the organization.
Thus, lateral phishers who wish to cast a wide phishing net might adopt a simple strategy of sending their attack to everyone at the organization.
We call these ATOs \emph{\atoOrgWide attackers} and identify them through two ways.

First, we search for any attackers where at least half of their phishing recipients belong to the ATO's organization,
and where at least 50\% of the organization's employees received the phishing email
(\ie the majority of a phisher's victims were employees and the attacker targeted a majority of the enterprise);
this search yielded a total of \numATOsOrgWideGeqHalfOrgRecip ATOs.
We estimate the list of an organization's employees by building a set of all employee email addresses
who sent or received email from anyone during the entire month of the phishing incident.\footnote{
This collection likely overestimates the actual set of employees because of service addresses, mailing list aliases,
and personnel churn.
}
For all of these \numATOsOrgWideGeqHalfOrgRecip ATOs,
less than \percentMaxContactBookOverlapOrgWideAgnostic of the recipients they target also appear in their recent contacts.
Coupled with the fact that each of these ATOs contacts over
\minNumRecipForOrgWideAttackersGeqHalfRecip recipients,
their behavior suggests that their initial goal focuses on phishing as many of the enterprise's recipients as possible,
rather than targeting users particularly close to the hijacked account.
Accordingly, we categorize them as \atoOrgWide attackers.

Our second heuristic looks for attackers whose recipient set consists nearly entirely of fellow employees,
but where the majority of the organization does not necessarily receive a phishing email.
Revisiting Figure~\ref{fig:recipDistribution}, \numATOsNearlyAllEmployeeRecip candidate \atoOrgWide ATOs
sent over 95\% of their phishing emails to fellow employee recipients.
However, we again need to exclude and account for ATOs who leverage their hijacked account's recent contacts.
From the first \atoOrgWide heuristic discussed previously,
we saw that less than \percentMaxContactBookOverlapOrgWideAgnostic of the recipients of that heuristic's \atoOrgWide attackers came from the ATO's recent contacts.
Using this value as a final threshold
for this second candidate set of \atoOrgWide attackers,
we identify \numATOsOrgWideNearlyAllEmployeeRecip \atoOrgWide attackers
where over 95\% of their recipients belong to the ATO's organization but less than
\percentMaxContactBookOverlapOrgWideAgnostic of the recipients came from the ATO's recent contacts;
a combination that suggests the attacker seeks primarily to compromise other employees,
but who do not necessarily have a personal connection with the hijacked account.

Aggregating and deduplicating the two sets of lateral phishers from above
produces a total of \numATOsOrgWide \atoOrgWide attackers (\percentATOsOrgWide), who
take advantage of the information in a hijacked account to target many fellow employees.

\begin{figure}
\includegraphics[width=1.0\columnwidth]{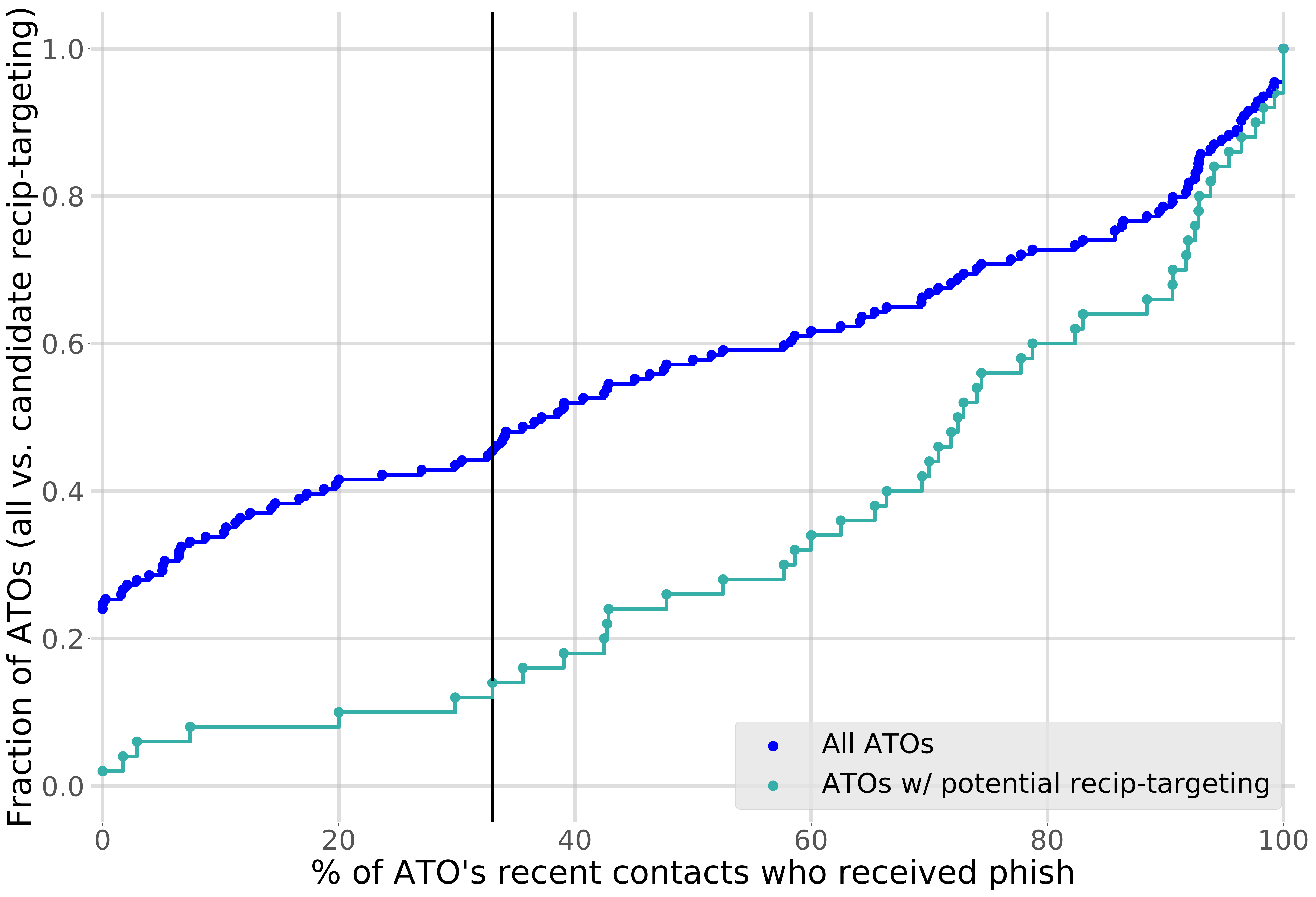}
\setlength{\abovecaptionskip}{-11pt}
\setlength{\belowcaptionskip}{-10pt}
\caption{
CDF: the $x$-axis displays what \% of the ATO's recent contacts received a lateral phishing email (\S~\ref{sec:targeting}).
The bottom teal graph filters the ATOs to \emph{exclude} any ATO identified as \atoRecipAgnostic, \atoLateralOrg, and \atoOrgWide attackers;
at the vertical black line, 88\% of these filtered ATOs send phishing emails to at least $x$ = 33\% of addresses from their recent contacts.
}
\label{fig:cdfContactBookPotentialRecipTargeting}
\end{figure}

\paragraph{\atoTargetedRecip Attackers}
For the remaining, uncategorized \numATOPotentialTargetedRecip ATOs,
we cannot conclusively determine the attackers' recipient targeting strategies
because our dataset does not provide us with the full set of information and actions available to the attacker.
Nonetheless, Figure~\ref{fig:cdfContactBookPotentialRecipTargeting}
presents some evidence that \numATOTargetedRecip of these remaining attackers do draw upon the hijacked account's prior relationships.
Specifically, \numATOTargetedRecip attackers sent their attacks to
at least 33\% of the addresses in the ATO's recent contacts.\footnote{
When examining and applying thresholds for the \atoRecipAgnostic and \atoOrgWide Attackers,
we used a slightly different fraction: how many of the ATO's phishing recipients also appeared in their recent contacts?
Here, we seek to capture attackers who make a specific effort to target a considerable number of familiar recipients.
Accordingly, we look at the fraction of the ATO's recent contacts that received phishing emails,
where the denominator reflects the number of users in the ATO's recent contacts,
rather than the ATO's total number of phishing recipients.
}
Since these ATOs sent attacks to at least 1 out of every 3 of the ATO's recently contacted recipients,
these attackers appear interested in targeting a substantial fraction of users with known ties to the hijacked account.
As such, we label these \numATOTargetedRecip ATOs as \atoTargetedRecip attackers.

\newcommand{\numPhishRecipOrgsFromNoisyATO}{14,835\xspace}
\newcommand{\numMajorityInternalRecipATO}{73\xspace}
\newcommand{\percentManyRecipDomainATO}{43\%\xspace}
\newcommand{\percentMajorityInternalRecipATO}{47.1\%\xspace}
\newcommand{\percentFamiliarRecipTargetingATO}{45\%\xspace}


\subsection{Message Content: Tailoring and Themes}\label{sec:messageContent}  
\newcommand{\numDistinctWords}{444\xspace}
Since lateral phishers control a legitimate employee account,
these attackers could easily mine recent emails to craft personalized spearphishing messages.
To understand how much attackers do leverage their privileged access in their phishing attacks,
this section characterizes the level of tailoring we see among lateral phishing messages.
Overall, only \percentIncidentsTargetedMsg of our dataset's incidents contain targeted content within their messages.
Across the phishing emails that used non-targeted content,
the attackers in our dataset
relied on two predominant narratives (deceptive pretexts) to lure their victim into performing a malicious action.
The combination of these two results suggests that, for the present moment,
these attackers (across dozens of organizations) see more value in opportunistically phishing as many recipients as possible,
rather than investing time to mine the hijacked accounts for personalized spearphishing fodder.

\paragraph{Content Tailoring}
When analyzing the phishing messages in our dataset,
we found that two dimensions aptly characterized the different levels of content tailoring and customization.
The first dimension, ``Topic tailoring'', describes how personalized the topic or main idea of the email is to the victim or organization.
The second dimension, ``Name tailoring'', describes how specifically the attacker addresses the victim (\eg ``Dear user'' vs. ``Dear Bob'').
For each of these two dimensions, we enumerate three different levels of tailoring and provide an anonymized message snippet below;
we use Bob to refer to one of the attack's recipients and FooCorp for the company that Bob works at.

\begin{enumerate}
  \item Topic tailoring: the uniqueness and relevancy of the message's topic to the victim or organization:
    \begin{enumerate}
      \item \label{contentTopicGeneric} Generic phishing topic: an unspecific message that could be sent to any user (``You have a new shared document available.'')
      \item \label{contentTopicEnterprise} Broadly enterprise related topic: a message that appears targeted to enterprise environments,
        but one that would also make sense if the attacker used it at many other organizations (``Updated work schedule. Please distribute to your teams.'')
      \item \label{contentTopicTargeted} Targeted topic: a message where the topic clearly relies on specific details about the recipient or organization
        (``Please see the attached announcement about FooCorp's 25th year anniversary.'', where FooCorp has existed for exactly 25 years.)
    \end{enumerate}
  \item Name tailoring: whether the phishing message specifically uses the recipient or organization's name:
    \begin{enumerate}
      \item \label{contentNameNone} Non-personalized naming: the attack does not mention the organization or recipient by name
        (``Dear user, we have detected an error in your mailbox settings...'')
      \item \label{contentNameOrg} Organization specifically named: the attack mentions just the organization, but not the recipient
        (``New secure email message from FooCorp...'')
      \item \label{contentNameRecip} Recipient specifically named: the attack specifically uses the victim's name in the email
        (``Bob, please review the attached purchase order...'')
    \end{enumerate}
\end{enumerate}

\begin{table}[t]
\begin{tabular}{@{}lrrr@{}}
\toprule
            & {\bf Generic} & {\bf Enterprise} & {\bf Targeted} \\
\midrule
No naming   & 90  & 35  & 9 \\
Organization named   & 23  & 16  & 4 \\
Recipient named & 0   & 3   & 0 \\
\bottomrule
\end{tabular}
\setlength{\belowcaptionskip}{-8pt}
\caption{Distribution of the number of \emph{incidents} per message tailoring category (\S~\ref{sec:messageContent}).
The columns correspond to how unique and specific the message's topic pertains to the victim or organization.
The rows correspond to whether the phishing email explicitly names the recipient or organization.
}
\label{table:contentTailoringStats}
\end{table}

Taken together, this taxonomy divides phishing content into nine different classes of tailoring;
Table~\ref{table:contentTailoringStats} shows how many of our dataset's \numIncidents incidents fall into each category.
From this categorization, two interesting observations emerge.
First, only \numIncidentsContentRecipNamed incidents (\percentIncidentsContentRecipNamed) actually address their recipients by name.
Since most ATOs (\percentATOsTotalRecipLarge) in our dataset email at least 100 recipients,
attackers would need to leverage some form of automation to both send hundreds of individual emails and customize the naming in each one.
Based on our results, it appears these attackers did not view that as a worthwhile investment.
For example, they might fear that sending many individual emails might trigger an anti-spam or anti-phishing mechanism,
which we observed in the case of one ATO who attempted to send hundreds of individual emails.
Second, looking at the last column of Table~\ref{table:contentTailoringStats},
only \numIncidentsContentTargeted incidents (\percentIncidentsTargetedMsg) use targeted content in their messages.
The overwhelming majority (\percentIncidentsContentNonTargeted) of incidents
opt for more generic messages that an attacker could deploy at a large number of organizations with minimal changes
(\eg by only changing the name of the victim organization).

While our attack dataset captures a limited view of all lateral phishing attacks,
it nonetheless reflects all known lateral phishing incidents across \numPhishOrgs organizations over a {\numMonths}-month timeframe.
Thus, despite the data's limitations, our results show that a substantial fraction of lateral phishers
do not fully draw upon their compromised account's resources (\ie historical emails) to craft personalized spearphishing messages.
This finding suggests these attackers act more like an opportunistic cybercriminal,
rather than an indomitable APT or nation-state.
However, given the arms-race and evolutionary nature of security,
these lateral phishers could in the future
increase the sophistication and potency of their attacks by drawing upon the account's prior emails to craft more targeted content.

\newcommand{\numRandomEmailsContentComparison}{1,000\xspace}
\newcommand{\numRandomConstitutedFromMalicious}{176\xspace}
\newcommand{\numRandomEmailTerms}{2,516\xspace}
\paragraph{Thematic Content (Lures)}
When labeling each phishing incident with a level of tailoring,
we noticed that the phishing messages in our dataset overwhelmingly relied on one of two deceptive pretexts (lures):
(1) an alarming message that asserts some problem with the recipient's account (and urges them to follow a link to remediate the issue);
and (2) a message that notifies the recipient of a new / updated / shared document.
For the latter `document' lure, the nature and specificity of the document varied with the level of content tailoring.
For example,
whereas an attack with generic topic tailoring will just mention a vague document,
attacks that use enterprise-related tailoring will switch the terminology to
an invoice, purchase order, or some other generic but work-related document.

\begin{table}
\centering
\begin{tabular}{@{}lr@{\hskip 0.25in}@{\hskip 0.25in}lr@{}}
\toprule
{\bf Word} & {\bf \# Incidents} & {\bf Word} & {\bf \# Incidents} \\ \midrule
 document & 89 &  sent & 44\\
 view & 76 &  review & 43  \\
 attach & 56 & share & 37 \\
 click & 55 &  account & 36 \\
 sign & 50  &  access & 34 \\
\bottomrule
\end{tabular}
\setlength{\belowcaptionskip}{-8pt}
\caption{Top 10 most common words across all \numIncidents lateral phishing incidents.
}
\label{table:topPhishWords}
\end{table}

To characterize this behavior further, we computed the most frequently occurring words across our dataset's phishing messages.
First, we selected one phishing email per incident,
to prevent incidents with many identical emails from biasing (inflating) the popularity of their lures.
Next, we normalized the text of each email:
we removed auto-generated text (\eg user signatures), lowercased all words, removed punctuation, and discarded all non-common English words;
all of these can be done with open source libraries such as Talon~\cite{talon} and NLTK~\cite{nltk}.
Finally, we built a set of all words that occurred in any phishing email across our incidents and counted how many incidents each word appeared in.

Interestingly, our dataset's phishing messages draw on a relatively small pool of words:
there are just \numDistinctWords distinct, common English words across the texts of every phishing message in our dataset
(\ie every phishing email's text consists of an arrangement from this set of \numDistinctWords words).
In contrast, a random sample of \numRandomEmailsContentComparison emails from our dataset
contained a total of \numRandomEmailTerms distinct words, and
only \numRandomConstitutedFromMalicious of these emails consisted entirely of words from the phishing term set.

Beyond this small set of total words across lateral phishing emails,
all but one incident contained at least one of the top 20 words,
illustrating the reliance on the two major lures we identified.
Figure~\ref{fig:phishPerWord} in Appendix~\ref{sec:techreportfigures} shows the full occurrence distribution of each word.
Focusing on just the top ten words and the number incidents that use them (Table~\ref{table:topPhishWords}),
the dominance of these two thematic lures becomes apparent.
Words indicative of the ``shared document'' lure, such as `document', `view', `attach', and `review', each occur in over 23\% of incidents,
with the most popular (document) occurring in nearly half of all incidents.
Similarly, we also see many words from the account-related lure in the top ten: `access', `sign' (from `sign on'), and `account'.

Overall, while our dataset contains several instances of targeted phishing messages,
the majority of the lateral phishing emails we observe rely on more mundane lures that an attacker can reuse across multiple organizations with little effort.
The fact that we see this behavior recur across dozens of different organizations suggests either
the emergence of a new, yet accessible, form of enterprise phishing, or
an evolution in the way ``ordinary'' cybercriminals execute phishing attacks
(moving from external accounts that use clever spoofing to compromised, yet legitimate accounts).

\subsection{Temporal Aspects of Lateral Phishing}\label{sec:temporal}

\begin{figure}[t]
\includegraphics[width=1.0\columnwidth]{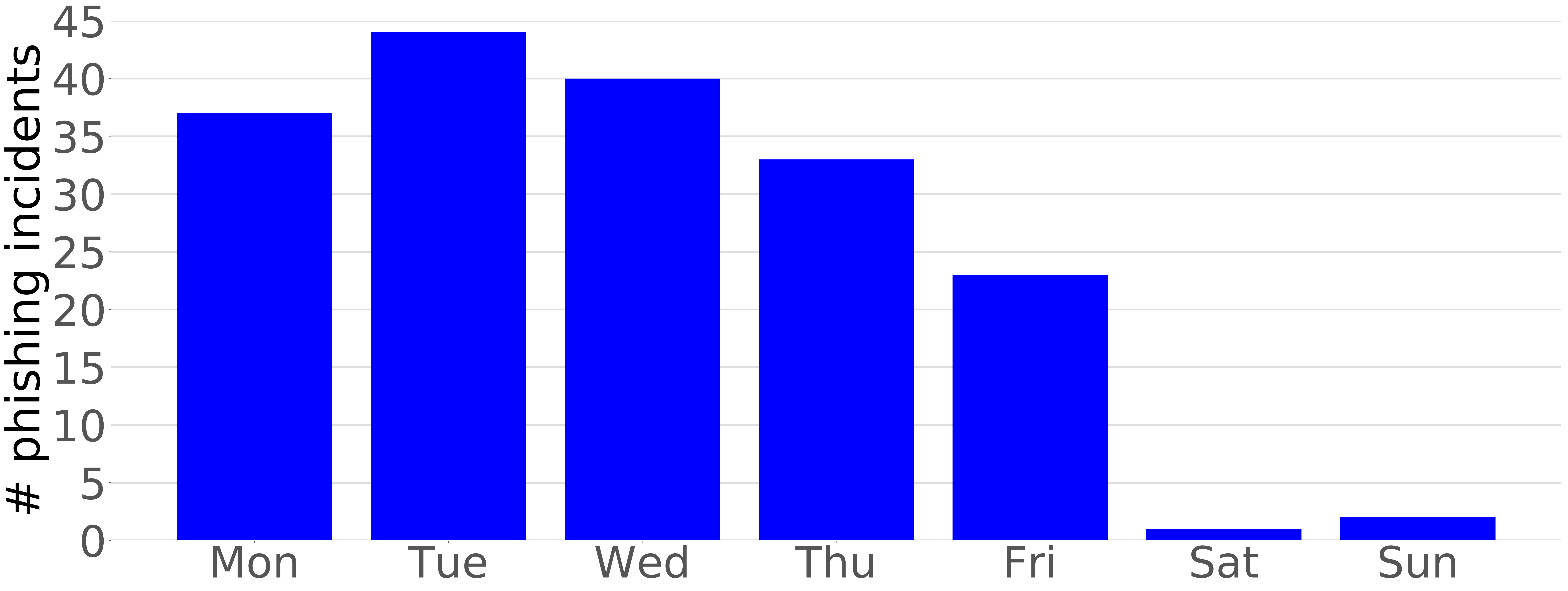}
\setlength{\abovecaptionskip}{-8pt}
\setlength{\belowcaptionskip}{-8pt}
\caption{Number of lateral phishing incidents per day of week.
}
\label{fig:incidentsPerDay}
\end{figure}

Because attackers might not live or operate in the same geographic region as the hijacked account,
prior work has suggested using features that capture unusual timing properties inherent in phishing emails~\cite{
stringhini2015ain, egele2013compa, gascon2018reading
}.
Contrary to this intuition, in our dataset
most lateral phishing attacks occur at ``normal'' times of the day and week.
First, for \percentIncidentsWorkDay of lateral phishing incidents,
the attacker sent the phishing email during a weekday.
Additionally, the majority of attackers in our dataset send their phishing emails during the true account's normal working hours.

\paragraph{Day of the Week}
From Figure~\ref{fig:incidentsPerDay}, all but three
lateral phishing incidents occurred during a work day (Monday--Friday).
This pattern suggests that attackers send their phishing emails on the same days when employees typically send their benign emails,
and that the day of the week will provide an ineffective or weak detection signal.
Moreover, \percentIncidentsFirstHalfOfWeek of incidents occur in the first half of the week (Mon--Wed),
indicating that the lateral phishers in our dataset do not follow
the folklore strategy where attackers favor launching their attacks on Friday
(hoping to capitalize on reduced security team operations over the
coming weekend)~\cite{attackerTimeFolklore}.

\paragraph{Time (Hour) of Day}
In addition to operating during the usual work week,
most attackers tend to send their lateral phishing emails during the typical working hours of their hijacked accounts.
To assess the (ab)normality of an attack's sent-time,
for each ATO, we gathered all of the emails that the account sent in the 30 days prior to their first lateral phishing email.
We then mapped the sent-time of each of these historical (and presumably benign) emails to the hour-of-day on a 24 hour scale,
thus forming a distribution of the typical hour-of-day in which each hijacked account usually sent their emails.
Finally, for each lateral phishing incident, we
computed the percentile for the phishing email's hour-of-day relative to the hour-of-day distribution for the ATO's historical emails.
For example,
phishing incidents with a percentile of 0 or 100 were sent at an earlier or later hour-of-day than any email that the
true account's owner sent in the preceding 30 days.

\begin{figure}[t]
\includegraphics[width=1.0\columnwidth]{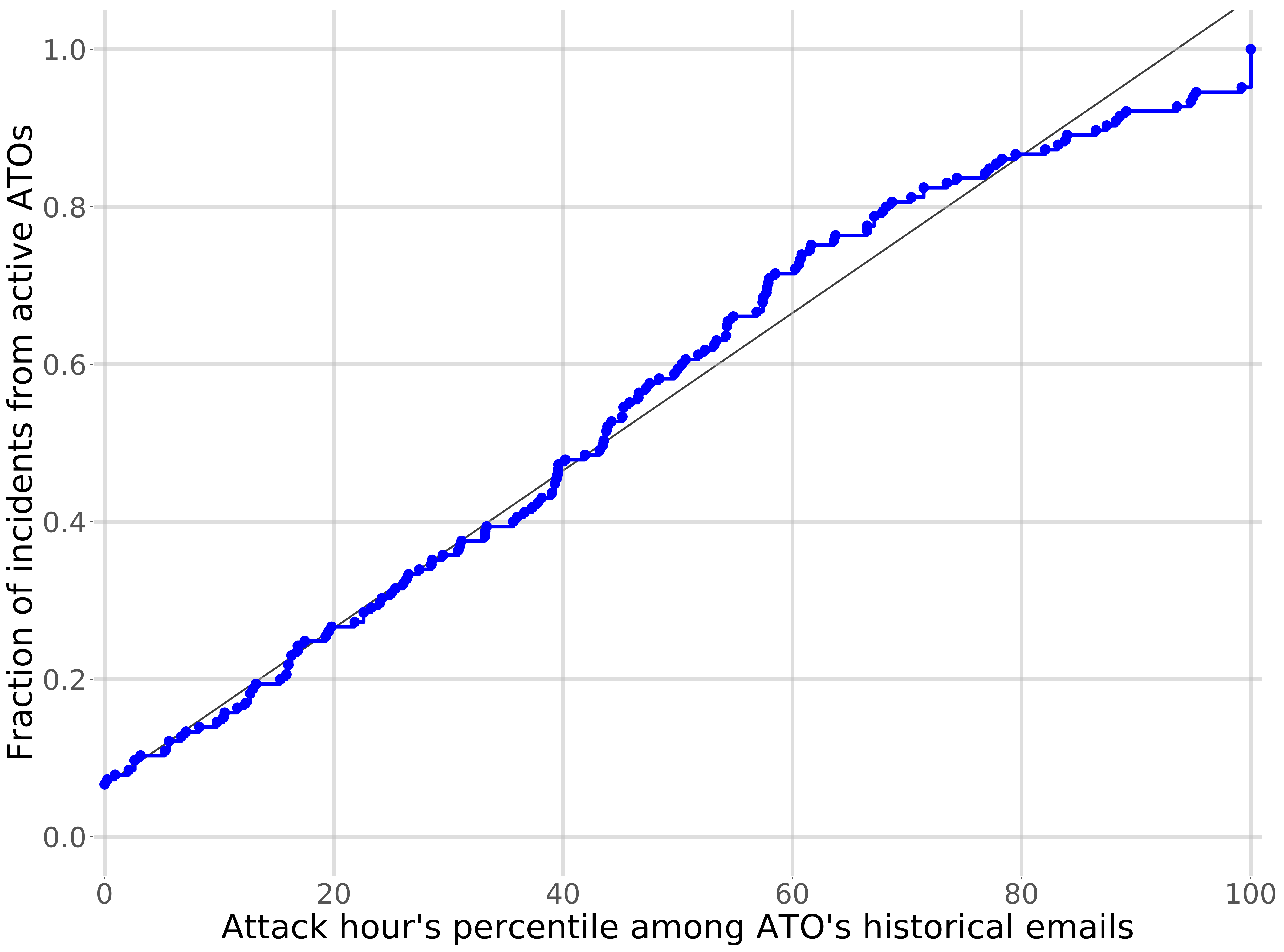}
\setlength{\abovecaptionskip}{-8pt}
\setlength{\belowcaptionskip}{-8pt}
\caption{CDF of the fraction of incidents from active ATOs where the time (hour) of day fell within the $x$'th percentile
of the hours at which the ATO's benign emails in the preceding 30 days were sent.
Active ATOs are hijacked accounts that sent at least 1 non-phishing email within the 30 days preceding their lateral phishing email.
}
\label{fig:cdfIncidentHourOfDayPercentiles}
\end{figure}  

Across all lateral phishing incidents sent by an active ATO,
Figure~\ref{fig:cdfIncidentHourOfDayPercentiles} shows what hour-of-day percentile the phishing incident's first email occurred at,
relative to the hijacked account's historical emails.
Out of the \numIncidents incidents, \numIncidentsQuiescentATO incidents were sent by an ``inactive'' (quiescent) ATO
that sent zero emails across all 30 days preceding their lateral phishing emails;
Figure~\ref{fig:cdfIncidentHourOfDayPercentiles} excludes these incidents.
Of the remaining \numIncidentsActiveATO incidents sent by an active ATO,
\numIncidentsOutsideNormalTime incidents fall completely outside of the hijacked account's historical operating hours,
which suggests that a feature looking for emails sent at atypical times for a user could help detect these attacks.
However, for the remaining \numIncidentsNormalTime incidents,
the phishing emails' hour-of-day evenly cover the full percentile range.
As shown in Figure~\ref{fig:cdfIncidentHourOfDayPercentiles},
the percentile distribution of phishing hours closely resembles the CDF of a uniformly random distribution
(a straight $y=x$ line);
i.e., the phishing email's hour-of-day appears to be randomly drawn from the true account's historical hour-of-day distribution.
This result indicates that for the majority of incidents in our dataset (\numIncidentsNormalTime out of \numIncidents),
the time of day when the ATO sent the attack will not provide a significant signal,
since their sent-times mirror the timing distribution of the true user's historical email activity.

Thus, based on the attacks in our dataset, we find that two weak timing-related features exist:
searching for quiescent accounts that suddenly begin to send suspicious emails (\numIncidentsQuiescentATO incidents),
and searching for suspicious emails sent completely outside of an account's historically active time window
(\numIncidentsOutsideNormalTime incidents).
Beyond these two features and the small fraction of phishing attacks they reflect,
neither the day of the week nor the time of day provide significant signals for detection.


\subsection{Attacker Sophistication}\label{sec:sophistication}
Since most of our dataset's lateral phishers do not mine the hijacked account's mailbox to craft targeted messages,
one might naturally conclude that these attackers are lazy or unsophisticated.
However, in this subsection, we identify two kinds of sophisticated behavior that required some investment of additional time and manual effort:
attackers who continually engage with their attack's recipients in an effort to increase the attack's success rate,
and attackers who actively ``clean up'' traces of their phishing activity in an attempt to hide their presence from the account's legitimate owner.
In contrast to the small number of attackers who invested time in crafting tailored phishing messages to a personalized set of recipients,
nearly one-third (\percentSophisticatedATO) of attackers engage in at least one of these two sophisticated behaviors.

\paragraph{Interaction with potential victims}
Upon receiving a phishing message,
some recipients naturally question the email's validity and send a reply asking for more information or assurances.
While a lazy attacker might ignore these recipients' replies,
\numInteractiveATO ATOs across \numOrgsWithInteractiveATO organizations actively engaged with these potential victims by sending follow-up messages assuring the victim of the phishing email's legitimacy.
For example, at one organization, an attacker consistently sent brief follow-up messages such as
``Yes I sent it to you'' or ``Yes, have you checked it yet?''.
In other cases, attackers replied with significantly more elaborate ruses:
\eg ``Hi [Bob], its a document about [X]. It's safe to open. You can view it by logging in with your email address and password.''

To find instances where a phisher actively followed-up with their attack's potential victims,
we gathered all of the messages in every lateral phishing email thread and checked
to see if the attacker ever received and responded to a recipient's reply (inquiry).\footnote{
Office 365 includes a \emph{ConversationID} field,
and all emails in the same thread (the original email and all replies) get assigned the same ConversationID value.
}
In total, we found that \numATOsWithInquiries ATOs received at least one reply from a recipient.
Of these reply-receiving attackers,
\numInteractiveATO ATOs (\percentInteractiveATO) sent a deceptive follow-up response to one or more of their recipients' inquiries.

\paragraph{Stealthiness}
Separate from interacting with their potential victims,
attackers might expend manual effort to hide their presence from the account's true owner
by removing any traces of their phishing emails,
particularly since lateral phishers appear to operate during the hijacked account's normal working hours (\S~\ref{sec:temporal}).
To estimate the number of these ATOs,
we searched for whether any of the following emails ended up in the hijacked account's Trash folder,
and were deleted within 30 seconds of being sent or received:
any phishing emails, replies to phishing emails, or follow-up emails sent by the attacker.
The 30 second threshold distinguishes stealthy behavior from deletion resulting from remediation of the compromised account.
In total, \numStealthyATO attackers across \numOrgsWithStealthyATO organizations engage in this kind of evasive clean-up behavior.

Of the \numInteractiveATO ATOs who interactively responded to inquiries about their attack,
only \numStealthyAndInteractiveATO also exhibited this stealthy clean-up behavior.
Thus, counting the number of attackers across both sets,
\numAnySophisticationATO ATOs engaged in at least one of these behaviors.

The sizeable fraction of attackers who engage in a sophisticated behavior creates a more complex picture of the attacks in our dataset.
Given that these attackers do invest dedicated and (often) manual effort in enhancing the success of their attacks,
why do so many of them (over 90\% in our dataset) use non-targeted phishing content and target dozens to hundreds of recipients?
One plausible reason for this generic behavior is that the simple methods they currently use work well enough under their economic model:
investing additional time to develop more tailored phishing emails just does not provide enough economic value.
Another reason might be that growth of lateral phishing attacks reflects an evolution in the space of phishing,
where previously ``simple'' external phishers have moved to sending their attacks via lateral phishing because attacks from (spoofed) external accounts have become too difficult,
due to user awareness and/or better technical mitigations against external phishing.
Ultimately, based on our work's dataset, we cannot soundly answer why so many lateral phishers employ simple attacks,
and leave it as an interesting question for future work to explore.

%% file: conclusion.tex
\section{Summary}\label{sec:conclusion}

In this work we presented the first large-scale characterization of lateral phishing attacks across
more than 100~million employee-sent emails from \totalOrgs enterprise organizations.
We also developed and evaluated a new detector that found many known lateral phishing attacks,
as well as dozens of unreported attacks, while generating a low volume of false positives.
Through a detailed analysis of the attacks in our dataset, we uncovered a number of important findings
that inform our mental models of the threats enterprises face,
and illuminate directions for future defenses.
Our work showed that \percentRandomOrgsWithPhish of our randomly sampled organizations, ranging from small to large,
experienced lateral phishing attacks within a seven-month time period,
and that attackers succeeded in compromising new accounts at least \percentSuccessfulATO of the time.
We uncovered and quantified several thematic recipient targeting strategies and deceptive content narratives;
while some attackers engage in targeted attacks,
most follow strategies that employ non-personalized phishing attacks that can be readily used across different organizations.
Despite this apparent lack of sophistication in tailoring and targeting their attacks,
\percentSophisticatedATO of our dataset's lateral phishers engaged in some form of sophisticated behavior
designed to increase their success rate or mask their presence from the hijacked account's true owner.
Additionally, over 80\% of attacks occurred during a typical working day and hour,
relative to the legitimate account's historical emailing behavior;
this suggests that these attackers either reside within a similar timezone as the accounts they hijack or make a concerted effort to operate during their victim's normal hours.
Ultimately, our work provides the first large-scale insights into an emerging, widespread form of enterprise phishing attacks,
and illuminates techniques and future ideas for defending against this potent threat.

%% file: acknowledgements.tex
\section*{Acknowledgements}
We thank Itay Bleier, the anonymous reviewers, and our shepherd Gianluca Stringhini for their valuable feedback.
This work was supported in part by the Hewlett Foundation through the
Center for Long-Term Cybersecurity, NSF grants CNS-1237265 and CNS-1705050,
an NSF GRFP Fellowship,
the Irwin Mark and Joan Klein Jacobs Chair in Information and Computer Science
(UCSD), by generous gifts from Google and Facebook, a Facebook Fellowship, and operational
support from the UCSD Center for Networked Systems.

%% file: appendix.tex
\appendix


\section{Detector Implementation and Evaluation Details}\label{sec:detectorImplementation}

\subsection{Labeling Phishing Emails}\label{sec:manualLabeling}

\paragraph{Labeling email as phishing or benign}
When manually labeling email, we started by examining five pieces of information: whether the email was a reported phishing incident,
the message content,
the suspicious URL that was flagged and if its domain made sense in context,
the email's recipients, and the sender.
With the exception of a few incidents,
we could easily identify a phishing email from the above steps.
For example: an email about a ``shared Office 365 document'' sent to hundreds of unrelated recipients and whose document link pointed to a bit.ly shortened (non-Microsoft) domain;
or an email describing an ``account security problem'' sent by a non-IT employee,
where the ``account reset'' URL pointed to an unrelated domain.
For the more difficult cases, we analyzed all replies and forwards in the email chain,
and labeled the email as phishing if it either received multiple replies / forwards that expressed alarm or suspicious,
or if the hijacked account eventually sent a reply saying that they did not send the phishing email.
Finally, as described in Section~\ref{sec:groundtruth},
we visited the non-side-effect, suspicious URLs from a sample of the labeled phishing emails.
All of the URLs we visited led to either an interstitial warning page (\eg Google SafeBrowsing),
or a spoofed log-on page.
For the emails flagged by our detector,
but which appeared benign based on examining all the above information,
we conservatively labeled them as false positives.
In many cases, false positives were readily apparent;
\eg emails where the ``suspicious URL'' flagged by our detector occurred in the sender's signature and linked to their personal website.

\paragraph{Training exercises vs. actual phishing emails}
In addition to distinguishing between a false positive and an attack,
we checked to ensure that our lateral phishing incidents represented actual attacks, and not training exercises.
First, based on the lateral phishing emails' headers, we verified that all of the sending accounts were legitimate enterprise accounts.
Second, all but five of the attack accounts sent one or more unrelated-to-phishing emails in the preceding month.
These two points gave us confidence that the phishing emails came from existing, legitimate accounts, and thus represented actual attacks;
\ie training exercises will not hijack an existing account, due to the potential reputational harm this could incur (and enterprise security teams we've previously engaged with do not do this).
Furthermore, none of our dataset's lateral phishing incidents are training exercises known to \cuda,
and none of the lateral phishing URLs used domains of known security companies.

\vfill\null

\subsection{Model Tuning and Hyperparameters}\label{sec:hyperparameters}
Most machine learning models, including Random Forest,
require the user to set various (hyper)parameters that govern the model's training process.
To determine the optimal set of hyperparameters for our classifier,
we followed machine learning best practices by conducting a three-fold cross-validation grid search over all combinations of the hyperparameters listed below~\cite{bergstra2012random}.
\begin{enumerate}
\item Number of trees: 50--500, in steps of 50 (\ie 50, 100, 150, \ldots, 450, 500)
\item Maximum tree depth: 10--100, in steps of 10
\item Minimum leaf size: 1, 2, 4, 8
\item Downsampling ratio of (benign / attack) emails: 10, 50, 100, 200
\end{enumerate}
Because our training dataset contained only a few dozen incidents,
we used three folds to ensure that each fold in the cross-validation contained several attack instances.
Our experiments used a Random Forest model with 64 trees, a maximum depth of 8, a minimum leaf size of 4 elements, and a downsampling of 200 benign emails per 1 attack email, since this configuration produced the highest AUC score~\cite{roccurve}.
But we note that many of the hyperparameter combinations yielded similar results.


\newpage

\section{Additional Detection Approaches}\label{sec:textSimilarityDetection}

In this section, we explore the two `suboptimal' text-based strategies that we alluded to earlier in Section~\ref{sec:design}.
First, we describe the features and classification process for each of these two strategies, and then evaluate their performance using the same time window and evaluation data from Section~\ref{sec:evaluation}.
Finally, we assess the performance of a detector that combines both these two additional strategies with the main strategy presented in Section~\ref{sec:design} and the overlap in detection across all three strategies.

\subsection{Design: Fuzzy Phish Matching Detector}\label{sec:detectorfuzzyphish}
The ``fuzzy phish matching'' detection strategy follows a natural intuition:
if the text in a new email closely matches the text in other known phishing emails,
then the new email is also probably a phishing attack.
In the context of the lure-exploit framework of phishing attacks~\cite{ho2017detecting},
this approach identifies phishing emails by looking for thematic lures that previously appeared,
and continue to recur, in new phishing attacks.

\paragraph{Features}
As input, this detector takes a set of known phishing emails.
Given a new email, the detector extracts a similarity feature, the email's \featurefuzzyphish, which measures the similarity between the new email's text and each known phishing email.

For each email, the detector normalizes the message's text by removing signatures and other auto-generated text with the Talon~\cite{talon} library, lowercasing all words, and removing punctuation.
Subsequently, the detector tokenizes the normalized text and converts it into a set of 3-grams of consecutive words.
To compute the similarity between two emails, we use the Jaccard similarity between these two sets of 3-grams.
After computing the pairwise similarity of the new email's text with all known phishing emails,
the detector assigns the highest similarity value as the email's \featurefuzzyphish.

To ensure a low volume of false positives,
we then combine this message-content feature with an additional feature that characterizes whether an email contains a potentially dangerous action.
Specifically, for each email, we extract a \featureglobalrep feature that quantifies the rarest URL an email contains
(using the same procedure discussed in~\S~\ref{sec:design}).

\paragraph{Classification}
Given a new email, we extract its \featurefuzzyphish and \featureglobalrep features.
Since our focus is on URL-based phishing, if an email contains no URLs,
we classify it as benign.
Otherwise, we apply a set of simple, conservative thresholds to these two features:
if an email's text is over 50\% similar to a known phish (its \featurefuzzyphish $\geq$ 50\%)
and it contains a domain whose global ranking is outside of the top 100,000 domains,
then we classify the email as phishing; otherwise, the detector labels it as benign.

\subsection{Design: Template Matching Detector}\label{sec:detectortemplate}
In a similar spirit to our previous strategy, our \detectorTemplate also tries to identify emails whose message content has a proclivity for phishing usage.
However, unlike the prior strategy,
which needed a set of historical phishing emails to recognize phishing text,
our \detectorTemplate attempts to automatically infer potential phishing content from a large corpus of mostly benign emails.
Since phishing emails frequently attempt to masquerade as a legitimate user or service,
this second detection approach attempts to build a set of \emph{template texts}:
``popular'' texts that users frequently encounter and associate with benign services.
For example, several attacks in our initial sample of lateral phishing emails contained messages that appeared nearly identical to the popular Docusign service.

\paragraph{Features}
We find these template texts by taking the past month's emails and then mapping each email to a tuple consisting of the email sender's domain and the alphabetically-ordered set of registered domains for all URLs embedded within the email (we call this alphabetically-ordered set the \emph{domain group} of an email).
Next, we keep only emails where both the email sender's domain and every domain in its domain group rank in the top 100,000 domains;
we also remove any email if the email sender's domain belongs to a popular email hosting provider (\eg Gmail, AOL, Comcast, etc.).
Additionally, we check that this (sender email domain, domain group) tuple appears in at least 50 emails per organization for at least 10\% of our organizations.
These requirements help ensure that we have a set of ``popular'' emails,
in that a reputable sender sent the email, all of its URLs belong to popular domains, and many different organizations receive this type of email.
Finally, to extract a set of ``popular texts'', we group all emails by their (sender email domain, domain group) value, and select
the email whose text has the highest 3-gram Jaccard similarity across all emails in the group.
That is, quantitatively, for each email, we compute its 3-gram Jaccard similarity
(as described earlier in~\S~\ref{sec:detectorfuzzyphish}) with every other email in the group.
We then sum up these similarity scores for each email, and select the email in each group with the highest summed score.
We refer to this resulting set of emails as a set of \emph{templates}.

Now that we have a set of templates, we can mimic the prior strategy.
First, we compute a \featuretemplatematch, which measures a new email's similarity to any known template,
by extracting 3-grams of consecutive words from the new email,
computing the Jaccard similarity of the email's 3-grams with every template's 3-grams,
and selecting the highest similarity score.
Finally, we extract the new email's \featureglobalrep feature using the same procedure as before (\S~\ref{sec:detectorfuzzyphish}).

\paragraph{Classification}
To classify a new email as lateral phishing or not, we apply the same thresholds to this detector's two features
as we did in our \detectorFuzzyPhish:
if a new email's text is over 50\% similar to any template, and the email contains a domain ranked outside of the top 100,000 domains,
our detector labels this new email as an attack; otherwise, it treats the email as benign.

\newcommand{\numAlertIncidentsTrainFuzzy}{2\xspace}
\newcommand{\numLateralIncidentsTrainFuzzy}{2\xspace}
\newcommand{\numFNIncidentsTrainFuzzy}{38\xspace}
\subsection{Evaluation: \detectorFuzzyPhish}\label{sec:evalfuzzy}
To evaluate our \detectorFuzzyPhish,
we aggregated all emails from \emph{user-reported} lateral phishing incidents,
and then used this set of phishing emails as our set of known-phishing emails for extracting a new email's \featurefuzzyphish (\S~\ref{sec:detectorfuzzyphish}).
To ensure temporal accuracy (\ie not using phish from the future to detect present phish),
we only compared a new email against known-phishing emails sent at least 24 hours prior.

\paragraph{Training and Tuning}
On our entire training dataset (\S~\ref{sec:evalmethodology}),
our \detectorFuzzyPhish generated alerts for \numAlertIncidentsTrainFuzzy
lateral phishing incidents (which were reported by users) and produces 0 false positives.

Despite generating no false alarms,
this approach missed \numFNIncidentsTrainFuzzy user-reported incidents.
These false negatives stem from two causes.
First, most known attacks in our dataset use very short email messages,
often consisting of a few words with an embedded phishing URL (\eg `New contract... View Here'),
followed by the hijacked user's signature.
Although we used several techniques to remove user signatures during our text similarity scoring,
at our scale of tens-of-millions of emails,
we encounter many email signatures which our set of techniques fail to remove.
This failure to remove signatures occurs for several of our known phishing emails,
causing them to generate poor text similarity scores with new phishing emails;
for short-message phishing emails,
the bulk of the text consists of the hijacked user's signature.
Second, very few phishing emails actually use the same or similar text.
For example, among phishing emails with short messages,
we observed many iterations of conceptually similar text that use different wording
(\eg `New contract' vs. `Alice shared Document X with you.')

\newcommand{\numAlertIncidentsTestFuzzy}{12\xspace}
\newcommand{\numLateralIncidentsTestFuzzy}{8\xspace}
\newcommand{\numNewLateralIncidentsTestFuzzy}{4\xspace}
\newcommand{\numFPIncidentsTestFuzzy}{4\xspace}
\newcommand{\numFNIncidentsTestFuzzy}{57\xspace}
\paragraph{Detection Results}
Turning to our test dataset (\S~\ref{sec:evalmethodology}),
this approach produced alerts for \numAlertIncidentsTestFuzzy incidents.
Of these, \numLateralIncidentsTestFuzzy are in fact lateral phishing;
the remaining  \numFPIncidentsTestFuzzy incidents are false positives.
For the same reasons we saw in our training dataset,
this detector exhibited a high false negative rate,
missing \numFNIncidentsTestFuzzy user-reported incidents.
Nonetheless, despite this strategy's high false negative rate,
we find that it generates virtually no false positives across a test dataset of tens-of-millions of emails.
Moreover, among the \numLateralIncidentsTestFuzzy lateral phishing incidents it detects,
\numNewLateralIncidentsTestFuzzy incidents were not reported by a user.

\newcommand{\numAlertIncidentsTrainTemplate}{5\xspace}
\newcommand{\numLateralIncidentsTrainTemplate}{4\xspace}
\newcommand{\numNewLateralIncidentsTrainTemplate}{2\xspace}
\newcommand{\numFNIncidentsTrainTemplate}{38\xspace}
\subsection{Evaluation: \detectorTemplate}\label{sec:evaltemplate}

\paragraph{Training and Tuning}
Prior to extracting features or classifying a new email,
our \detectorTemplate uses the past month's emails (across all of our training organizations)
to generate a set of templates (\S~\ref{sec:detectortemplate}).
Then, given a new email, this approach extracts the email's features and classifies it as described earlier in Section~\ref{sec:detectortemplate}.

Running this approach on our training dataset,
we find that this strategy correctly flagged \numLateralIncidentsTrainTemplate incidents as lateral phishing
(where~\numNewLateralIncidentsTrainTemplate incidents do not appear to be reported by a user).
It generated only one additional alert, which turned out to be a phishing training exercise email.
In all of these cases, the attackers (and simulation) closely mimicked a legitimate Docusign email's content,
but replaced the main `shared document' link with a phishing URL.

Examining the user-reported incidents it missed (\numFNIncidentsTrainTemplate),
we find that the \detectorTemplate is simply ill-suited for identifying the majority of lateral phishing emails:
the text of the phish it missed simply does not appear to mimic any popular, real email.
For example, for many of the training dataset attacks it failed to detect,
the phishing emails often presented only a short message such as ``Please see attached'' or ``Here is the new document'' (in addition to the hijacked user's signature).

\newcommand{\numAlertIncidentsTestTemplate}{8\xspace}
\paragraph{Detection Results}
Across our test dataset, our \detectorTemplate generated alerts for only \numAlertIncidentsTestTemplate total incidents,
all of which are phishing emails that come from external sources who spoof a fake username at the victim organization.
Mirroring our training dataset findings,
all of the user-reported incidents in our test dataset contain phishing messages which do not closely match a legitimate, popular email's text.
As such, based on the underlying motivations and assumptions for our \detectorTemplate,
this approach will be unlikely to detect these kinds of attacks.
Nonetheless, as with our training dataset,
this detection strategy produced 0 false positives across tens of millions of emails,
while still flagging several phishing incidents (albeit ones generated by external spoofing).

\subsection{Combined Detector}\label{sec:combinedDetector}
We can combine our three detection strategies (including our main approach~\S~\ref{sec:design}) into one detector by labeling a new email as lateral phishing if any of the techniques deems it phishing;
throughout this section, we also refer to each of the strategies as a `subdetector'.

\newcommand{\recallAggregateTestForTrainingOrgs}{81.8\%\xspace}
\newcommand{\recallAggregateTestForTestOrgs}{91.0\%\xspace}
\newcommand{\precisionAggregateTestForTrainingOrgs}{24.8\%\xspace}
\newcommand{\precisionAggregateTestForTestOrgs}{22.8\%\xspace}

\paragraph{Combined Detection Results}
On the entire test dataset, this aggregate detector achieved a Recall of~\recallAggregateTest,
a Precision of~\precisionAggregateTest, and
a False Positive Rate of~\fpPercentAggregateTest (one false positive per~\aggregateDetectorNumBenignPerFP employee-sent emails).
Although its precision is lower than desired,
the overall low volume of false alarms it generates could enable this detector
to be operationally viable.

Our test dataset consists of \numTrainingOrgs organizations from our training dataset,
plus a held-out set of \numTestOrgs new organizations;
we find that our detector performed comparably on both.
Breaking down our test dataset performance numbers, our detector achieved a recall of
\recallAggregateTestForTrainingOrgs and a precision of \precisionAggregateTestForTrainingOrgs across our training organizations versus
a recall of \recallAggregateTestForTestOrgs and a precision of \precisionAggregateTestForTestOrgs for the held-out orgs.

\newcommand{\numPhishDetectedInEvalTimeframe}{97\xspace}
\newcommand{\numPhishDetectedInEvalTimeframeOneDetector}{90\xspace}
\newcommand{\numPhishDetectedInEvalTimeframeTwoDetectors}{7\xspace}
\paragraph{Detector overlap}
Of the \numPhishDetectedInEvalTimeframe test dataset incidents found by our aggregate detector,
\numPhishDetectedInEvalTimeframeOneDetector are detected by only one subdetector,
and \numPhishDetectedInEvalTimeframeTwoDetectors incidents are detected by two subdetectors.
For all of the \numPhishDetectedInEvalTimeframeTwoDetectors incidents,
the two overlapping subdetectors that detected them are the \detectorFuzzyPhish and the primary detection strategy (\S~\ref{sec:design}).
Of the remaining \numPhishDetectedInEvalTimeframeOneDetector incidents,
the primary subdetector is responsible for all but one incident's detection,
with the final one coming from the \detectorFuzzyPhish.
As we explored earlier in Section~\ref{sec:evalfuzzy},
this result reflects the fact that the text of phishing emails exhibits frequent churn over time,
causing our two text-similarity driven strategies to miss new attacks that our main approach detects.

\newpage

\section{Additional Figures}\label{sec:techreportfigures}

\begin{figure}[h]
\includegraphics[width=1.0\columnwidth]{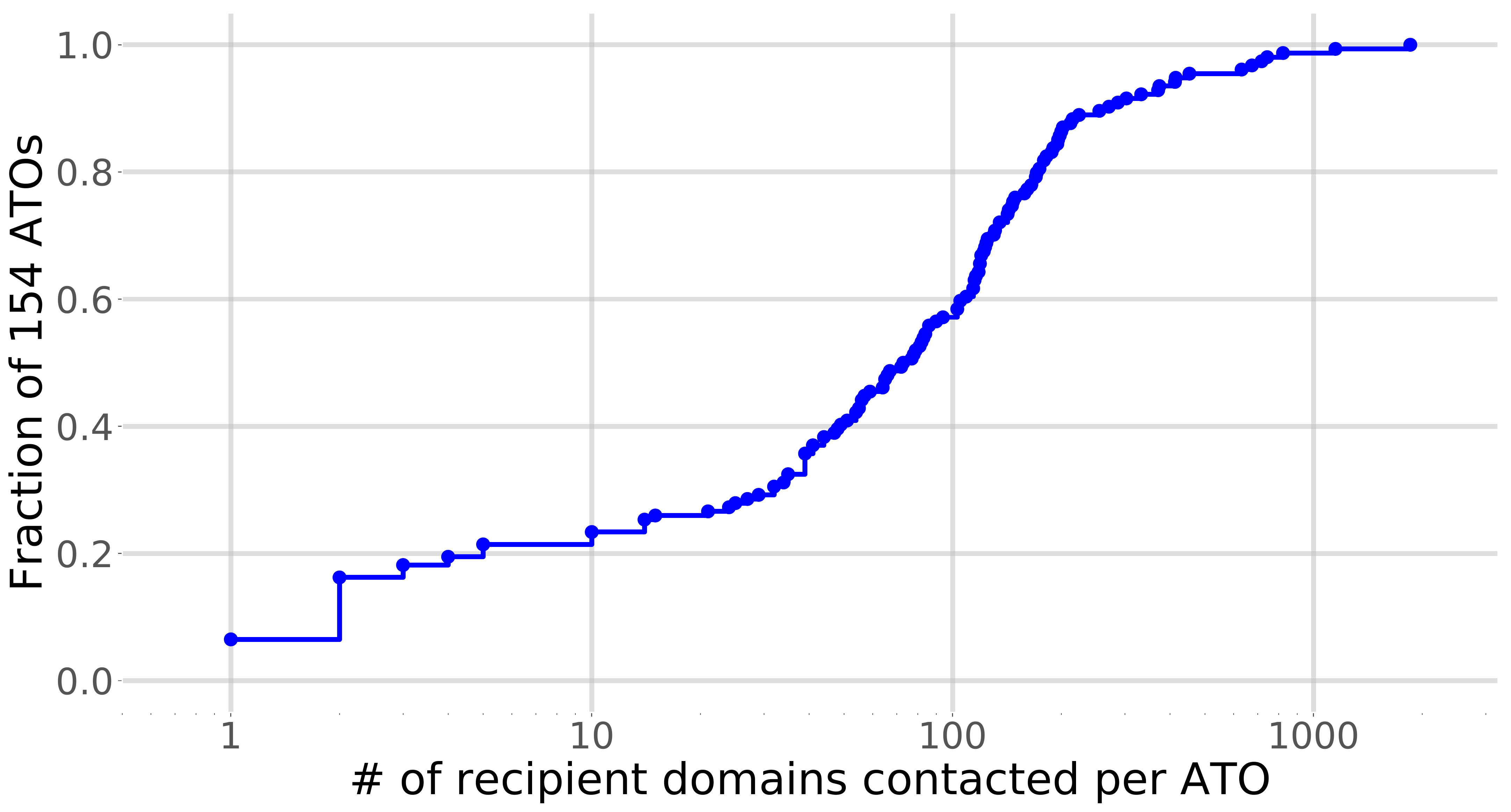}
\caption{CDF of the fraction of lateral phishers who send attacks to $x$ distinct recipient email domains.
(\S~\ref{sec:targeting})}
\label{fig:recipDomainsPerATO}
\end{figure}

\paragraph{Reach of lateral phishing attacks}
Figure~\ref{fig:recipDomainsPerATO} shows an approximation of the number of
different organizations (recipient address domains) targeted by the attackers in our dataset.
Nearly 80\% of lateral phishers in our dataset send their attacks to recipients
at 10 or more organizations.

\begin{figure}[h]
\includegraphics[width=1.0\columnwidth]{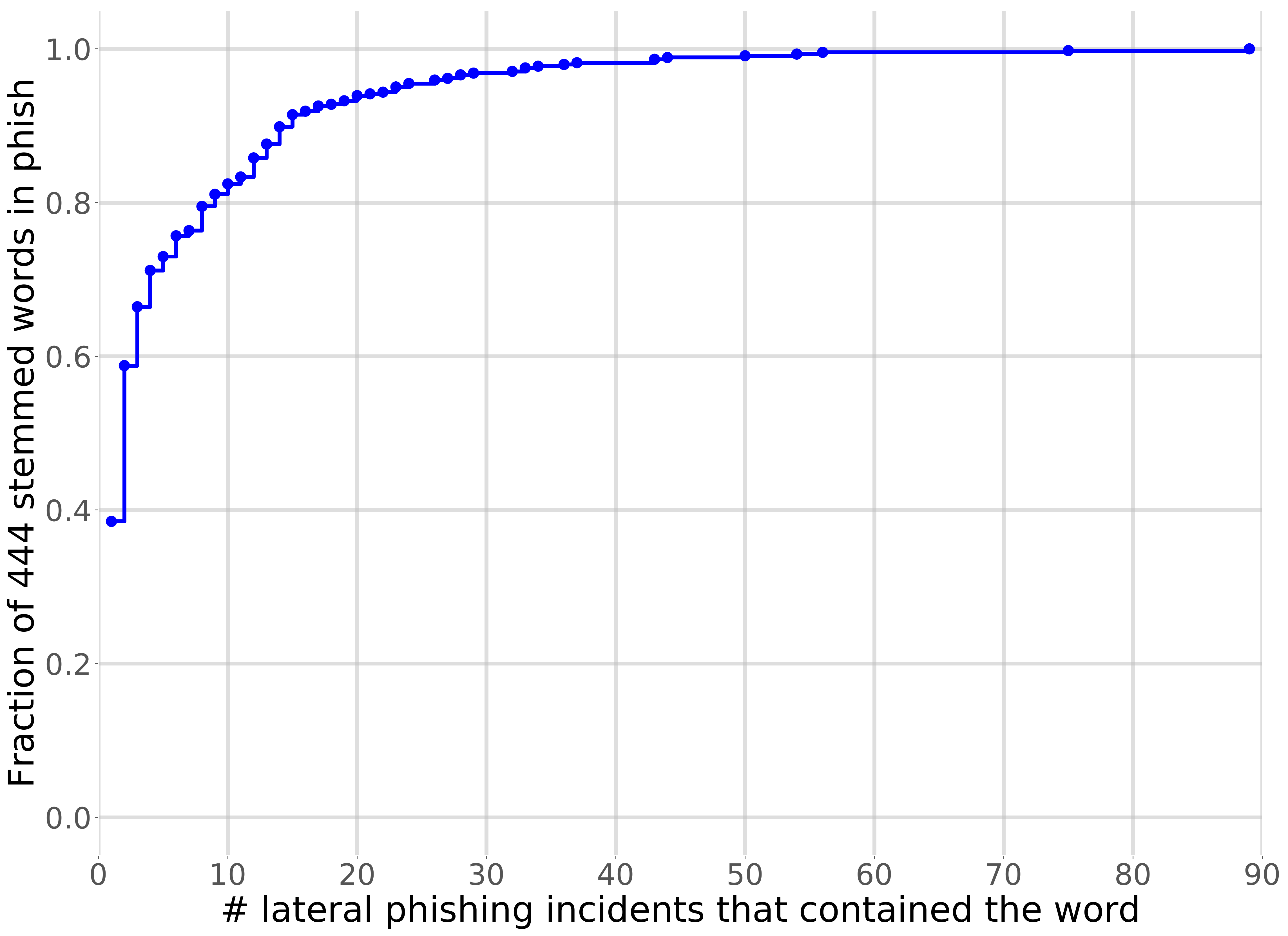}
\caption{
CDF of the fraction of common, stemmed English words that appear in $x$ lateral phishing incidents. (\S~\ref{sec:messageContent})}
\label{fig:phishPerWord}
\end{figure}

\paragraph{Language of lateral phishing messages}
Figure~\ref{fig:phishPerWord} shows the distribution for how often a word appeared across our dataset's lateral phishing incidents.
Earlier in Section~\ref{sec:messageContent}, Table~\ref{table:topPhishWords}
showed the frequencies for the top 10 words across lateral phishing incidents.

\newpage

\begin{figure}[h]
\includegraphics[width=1.0\columnwidth]{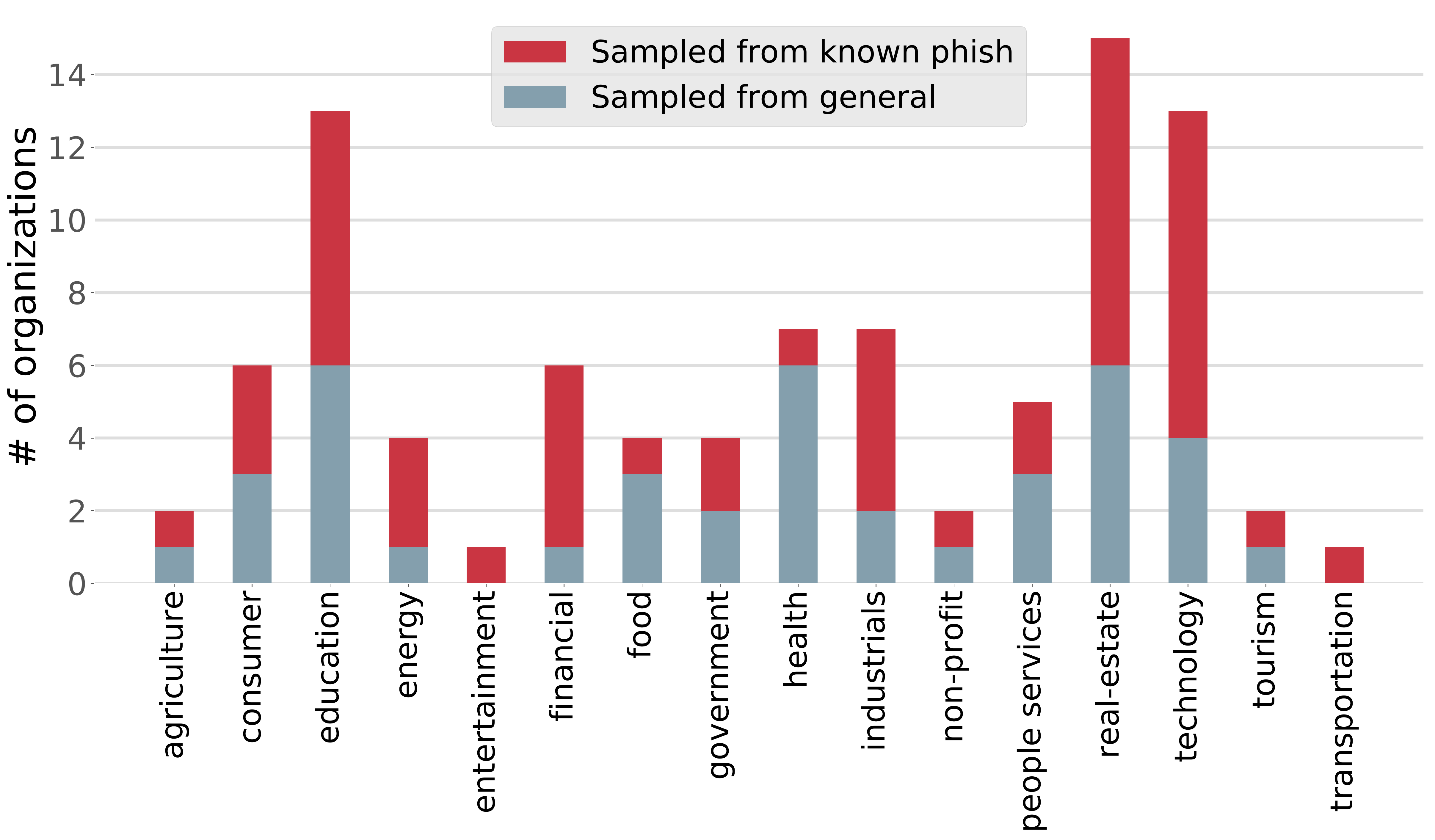}
\caption{Distribution of our dataset's organizations by economic sector:
organizations that came from sampling the pool of organizations with reported phish versus
the ones that came from sampling the pool of all organizations.}
\label{fig:orgSectorsBySamplingMethod}
\end{figure}

\begin{figure}[h]
\includegraphics[width=1.0\columnwidth]{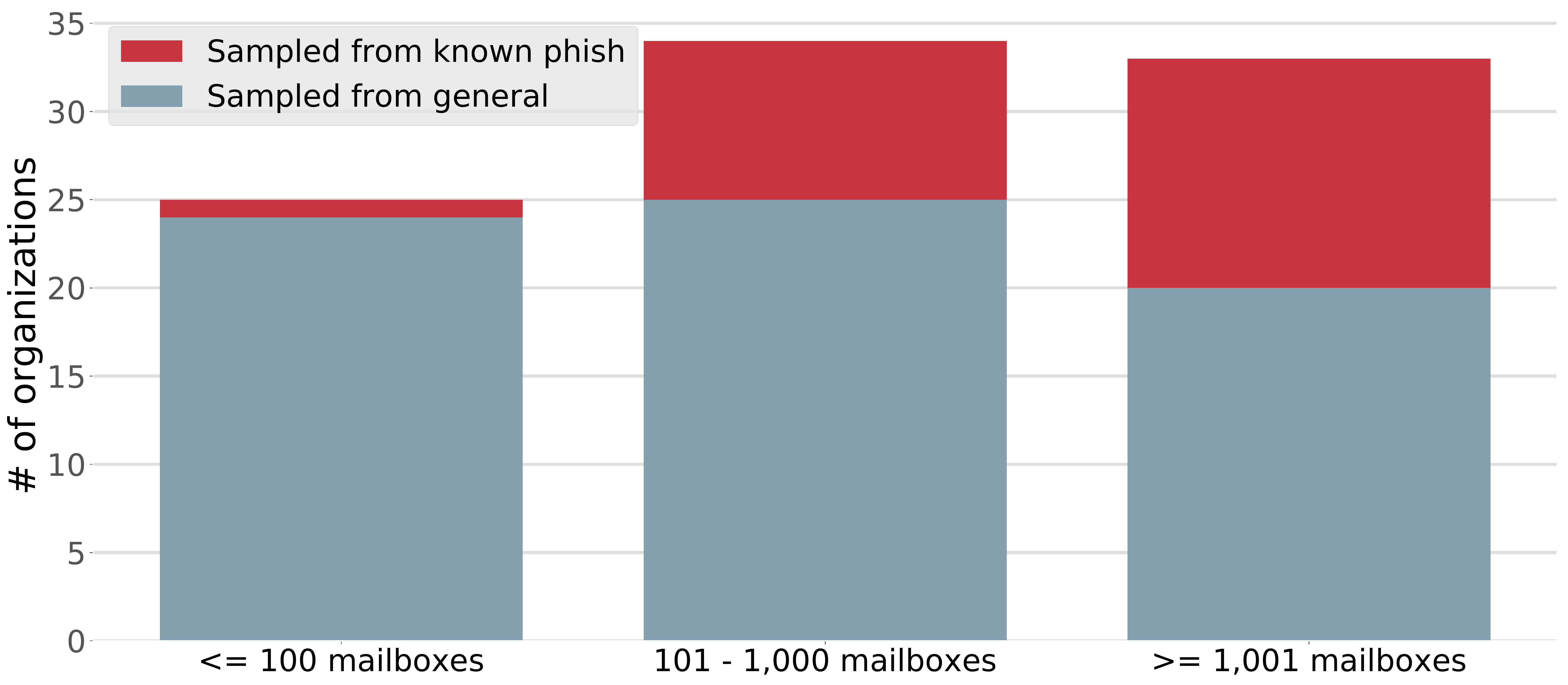}
\caption{Distribution of our dataset's organizations by size (number of accounts):
organizations that came from sampling the pool of organizations with reported phish versus
the ones that came from sampling the pool of all organizations.
}
\label{fig:orgSizesBySamplingMethod}
\end{figure}

\paragraph{Distribution of organizations by sampling method}
Figures~\ref{fig:orgSectorsBySamplingMethod} and~\ref{fig:orgSizesBySamplingMethod}
show further details about the organizations in our dataset, extending
the characterization in Section~\ref{sec:dataset}.
In particular, Figure~\ref{fig:orgSectorsBySamplingMethod}
shows the economic sector
and Figure~\ref{fig:orgSizesBySamplingMethod} shows the size distributions of our dataset's organizations
by sampling method
(\ie the organizations we sampled from those with reported lateral phishing
versus those we sampled from all organizations).

